\documentclass[aps,prx,longbibliography,a4paper,notitlepage,reprint,nofootinbib,
]{revtex4-2}

\usepackage{amsmath,amssymb,amsfonts} 
\usepackage{mathrsfs} 
\usepackage{mathtools} 
\usepackage{color}
\usepackage[usenames,dvipsnames]{xcolor}
\usepackage{graphicx,float}
\usepackage{hyperref}
\hypersetup{colorlinks,linkcolor=blue,citecolor=blue,urlcolor=blue}
\usepackage{cleveref}
\usepackage{soul}
\usepackage{siunitx}

\usepackage[switch]{lineno} 
\usepackage{lipsum}

\makeatletter
\renewcommand\frontmatter@abstractwidth{\dimexpr\textwidth\relax}
\makeatother

\makeatletter
\newcommand{\fmarki}{*}
\newcommand{\fmarkii}{\ensuremath{\dagger}}
\def\@fnsymbol#1{{\ifcase#1\or \fmarki\or \fmarkii \else\@ctrerr\fi}}
\makeatother

\renewcommand{\fmarki}{$\dagger$}
\renewcommand{\fmarkii}{$\ddagger$}

\begin{document}
\title{
Coherent optical control of a superconducting microwave cavity\\ via electro-optical dynamical  back-action 
}
\author{Liu Qiu$^\star$}
\email{liu.qiu@ist.ac.at}
\author{Rishabh Sahu$^\star$}
\author{William Hease}
\author{Georg Arnold}
\author{Johannes M. Fink}
\email{jfink@ist.ac.at}
\affiliation{Institute of Science and Technology Austria, Am Campus 1, 3400 Klosterneuburg, Austria}

\date{\today}

\begin{abstract}
\noindent 
Recent quantum technologies have established precise quantum control of various microscopic systems using electromagnetic waves.
Interfaces based on cryogenic cavity electro-optic systems are particularly promising, due to the direct 
interaction between microwave and optical fields in the quantum regime. 
Quantum optical control of superconducting microwave circuits has been precluded so far due to the weak electro-optical coupling as well as  quasi-particles induced by the pump laser. 
Here we report the coherent control of a superconducting microwave cavity using laser pulses in a multimode electro-optical device at millikelvin temperature with near-unity cooperativity.
Both the stationary and instantaneous responses of the microwave and optical modes comply with the coherent electro-optical 
interaction, and reveal only minuscule amount of excess back-action with an unanticipated time delay. 
Our demonstration enables wide ranges of applications
beyond quantum transductions, from squeezing and quantum non-demolition measurements of microwave fields, to entanglement generation and hybrid quantum networks.
\end{abstract}

\maketitle
\def\thefootnote{$\star$}
\footnotetext{These authors contributed equally to this work.}
\def\thefootnote{\arabic{footnote}}

Microwave superconducting quantum technologies have facilitated the electronic readout and control of superconducting circuits and quantum dot spin qubits~\cite{blais_circuit_2021,petta_coherent_2005}, which holds the promise for quantum-enhanced sensing~\cite{lloyd_enhanced_2008} and scalable quantum computing~\cite{arute_quantum_2019}.
Emerging challenges include interfacing the superconducting circuits to complex electrical lines, which introduces excess heat load and complexity beyond traditional cryogenic systems.
Photonic fiber links, due to the low propagation loss and passive heating, can be adopted to deliver microwave signals for quantum circuits readout and control at millikelvin temperatures, e.g. using photodiodes~\cite{lecocq_control_2021}, mechanical transducers~\cite{mirhosseini_superconducting_2020,delaney_superconductingqubit_2022}, or microwave photonics~\cite{marpaung_integrated_2019, youssefi_cryogenic_2021}.
Despite the ubiquitous electro-optic  devices in modern telecommunication networks with ultra-high speed translation between electronic and optical fields~\cite{maleki_optoelectronic_2011,reed_silicon_2010,wang_integrated_2018}, their operations in the quantum regime have been impeded so far due to the weak electro-optical coupling, even at cryogenic temperatures~\cite{youssefi_cryogenic_2021}.

Cavity electro-optics (CEO) employs resonantly-enhanced electro-optic interaction with optimized spatial overlap of microwave and optical modes~\cite{ilchenko_whisperinggallerymode_2003,tsang_cavity_2010}.
It holds great promises for general quantum measurement and control of superconducting microwave circuits with optical laser light~\cite{tsang_cavity_2010,tsang_cavity_2011,rueda_efficient_2016,soltani_efficient_2017}, ranging from microwave-optical entanglement generation~\cite{rueda_electrooptic_2019,matsko_fundamental_2007,sahu_2022}, coherent microwave or optical signal synthesis~\cite{tsang_cavity_2010},
to laser cooling of the microwave mode~\cite{sahu_quantumenabled_2022}, 
and bidirectional microwave-optical quantum transduction with near unity efficiency and low added noise~\cite{fan_superconducting_2018,hease_bidirectional_2020,xu_bidirectional_2021,sahu_quantumenabled_2022}.
A multimode CEO system allows for quantum thermometry~\cite{scigliuzzo_primary_2020,weinstein_observation_2014a} and quantum non-demolition measurements of the microwave field beyond the standard quantum limit with significantly reduced probing powers~\cite{braginsky_quantum_1980,shomroni_optical_2019, matsko_fundamental_2007,dobrindt_theoretical_2010,kronwald_arbitrarily_2013a}.
One particularly promising application of CEO is to build a complex optical quantum network connecting hybrid superconducting microwave quantum circuits~\cite{wehner_quantum_2018,clerk_hybrid_2020}, 
with alternative approaches using electro- or piezo-optomechanical devices~\cite{andrews_bidirectional_2014b,delaney_superconductingqubit_2022, bochmann_nanomechanical_2013b,mirhosseini_superconducting_2020}, trapped atoms~\cite{tu_highefficiency_2022,kumar_quantumlimited_2022}, rare-earth ions doped crystals ~\cite{williamson_magnetooptic_2014} and optomagnonic devices~\cite{zhu_waveguide_2020,han_microwaveoptical_2021}.

Such prospects rely on the optical coherent dynamical control of the superconducting microwave cavity, i.e. via the electro-optical dynamical back-action (DBA)~\cite{tsang_cavity_2010}. This has been impeded so far due to the typically weak electro-optical coupling, or the significant excess back-action, i.e. unwanted perturbations that are not due to the electro-optic effect,
as a result of the required strong optical pump.
Despite the steady progress in the last years, primarily on quantum transductions~\cite{fan_superconducting_2018,hease_bidirectional_2020,xu_bidirectional_2021,sahu_quantumenabled_2022}, most CEO systems suffer from limited cooperativity $C$~\cite{tsang_cavity_2010, javerzac-galy_onchip_2016}, a measure for coherent coupling versus the microwave and optical dissipation.
An endeavor towards coherent electro-optical interaction at unitary cooperativity has started in the last years, including explorations in various electro-optic materials and fabrication processes, e.g. based on 
aluminum nitride
~\cite{fan_superconducting_2018,fu_cavity_2021}, 
bulk and thin-film lithium niobate (LN)
~\cite{rueda_efficient_2016,hease_bidirectional_2020,xu_bidirectional_2021,mckenna_cryogenic_2020,holzgrafe_cavity_2020,sahu_quantumenabled_2022},
barium titanate~\cite{eltes_integrated_2020} 
and organic polymers~\cite{witmer_siliconorganic_2020}. 
However, excess  dissipation~\cite{zhang_monolithic_2017,zhu_integrated_2021} and  back-action  still remain in optical and microwave resonators, originating from, e.g. piezoelectric~\cite{holzgrafe_cavity_2020,mckenna_cryogenic_2020}, photorefractive effects~\cite{xu_photorefractioninduced_2021,xu_mitigating_2021}, absorption~\cite{zhu_integrated_2021}, dissipative feedback~\cite{qiu_dissipative_2022}, quasi-particles~\cite{witmer_siliconorganic_2020,xu_lightinduced_2022}, etc.

Pulsed operation in CEO devices reduces the integrated optical power while maintaining the  cooperativity, and has recently enabled demonstrations of quantum transduction in the microwave ground state~\cite{fu_cavity_2021,sahu_quantumenabled_2022}.
The compatibility of CEO devices to superconducting microwave circuits calls for resolving and controlling pulsed microwave signals in the time domain in a nondestructive manner~\cite{mirhosseini_superconducting_2020, arute_quantum_2019,lecocq_control_2021,delaney_superconductingqubit_2022}.
However, the coherent optical dynamical control of superconducting microwave cavity has remained elusive.

In this work we demonstrate coherent electro-optic dynamical back-action in a multimode cavity electro-optic device.
Our results demonstrate coherent stationary and instantaneous electro-optic DBA to the microwave mode,
such as the optical spring effect and microwave linewidth narrowing or broadening, with negligible excess back-action.
We  observe electro-optically induced absorption or transparency  of the optical probing field~\cite{boller_observation_1991,weis_optomechanically_2010,safavi-naeini_electromagnetically_2011,fan_superconducting_2018}, which opens up the possibility for dispersion engineering of propagating optical and microwave pulses. 
The observed coherent electro-optical response confirms the feasibility of our multimode CEO system for the direct quantum optical control and sensing of microwave fields in the quantum back-action (QBA) dominant regime~\cite{tsang_cavity_2010}, and provides important insights into the complex time-dependence of pulsed quantum protocols, e.g. electro-optic entanglement generation~\cite{sahu_2022}.

\begin{figure*}[t]
	\includegraphics[scale=1]{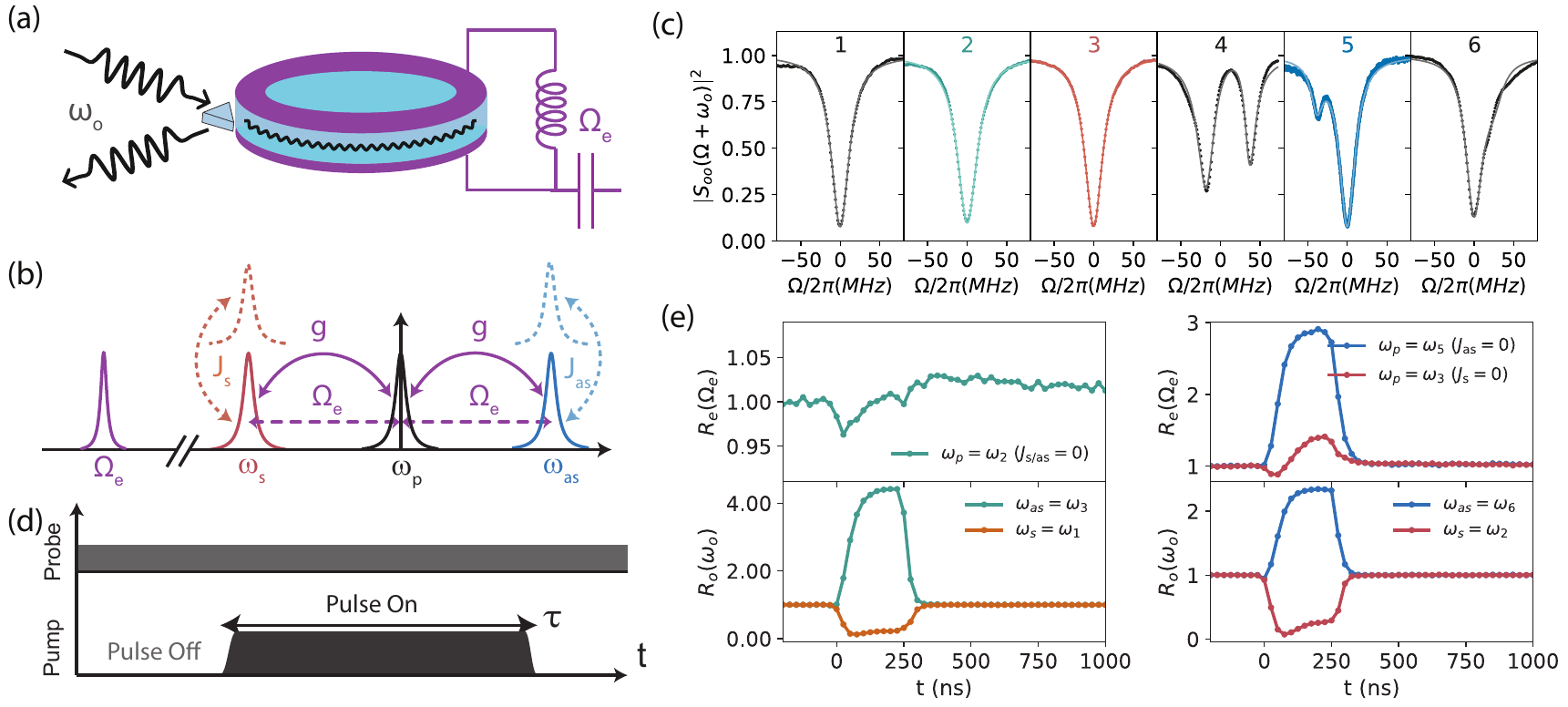}
	\caption{\textbf{Multimode cavity electro-optical system in the pulsed regime.} 
		\textbf{a}, Schematic representation of the cavity electro-optic device. 
		A millimeter-sized lithium niobate
		optical resonator (light blue) is placed in the capacitor of the LC circuit realized as an aluminum 3D microwave cavity (purple). 
		Optical light is fed to the EO device via an antireflection-coated diamond prism.
		\textbf{b},  Mode configurations of the CEO device, with one microwave mode (purple) coupled to three optical TE modes, i.e.~the Stokes (red), pump (black) and anti-Stokes mode (blue).
		A strong optical pump of frequency $\omega_p$ generates pump enhanced Stokes and anti-Stokes scattering at a rate $g$, which can be selectively suppressed by coupling to an optical TM mode (dashed curve).  
		\textbf{c}, Measured optical reflection (dots) of a series of modes with fitting curves (lines), with mode 2 as the pump mode for the symmetric case ($J_{\mathrm{s/as}} =0$)
		while mode 3 and 5 for the Stokes ($J_{s}=0$) and anti-Stokes ($J_{as}=0$) case respectively. 
		All resonances are re-centered to the individual TE mode resonance frequency.
		Mode splitting in mode 4 indicates strong TE-TM mode coupling.
		\textbf{d}, Coherent dynamical response probing scheme, with a short optical pump pulse of duration $\tau$ and a weak continuous probing field around the microwave or optical (Stokes or anti-Stokes) mode frequency.
		\textbf{e}, Temporal on-resonance response $R(\omega)$[cf. Eq.~\ref{eq:R}], i.e.~the normalized probing field reflection  between pulse on and off (pump peak power $\sim 500\, \mathrm{mW}$).
		Left panel shows the symmetric case ($\omega_p = \omega_2$), with on-resonance microwave response (green curve) in the upper panel and optical Stokes ($\omega_{s} = \omega_1$, orange curve) and anti-Stokes ($\omega_{as} = \omega_3$, green curve) responses in the lower panel.
		Right panel shows the two asymmetric cases, i.e.~the on-resonance microwave and optical Stokes responses ($\omega_{s} = \omega_2$) in the Stokes case ($\omega_p = \omega_3$, red curves), and the on-resonance microwave and optical anti-Stoke responses ($\omega_{as} = \omega_6$) in the anti-Stokes case ($\omega_p = \omega_5$, blue curves).
		}\label{fig:1}
\end{figure*}

\vspace{0.25cm}
\noindent
\textbf{Results}\\
\noindent
\textbf{Theoretical Model and Experiment}
We realize this experiment in a multimode cavity electro-optical device~\cite{rueda_efficient_2016} as depicted in Fig.~\ref{fig:1}(a), where a crystalline lithium niobate whispering gallery mode (WGM) optical resonator is coupled to the azimuthal number $m=1$ mode of a superconducting aluminum microwave cavity inside a dilution refrigerator at~$\sim$10~mK \cite{hease_bidirectional_2020,sahu_quantumenabled_2022}.
As shown in Fig.~\ref{fig:1}(b), we consider a series of optical transverse-electric (TE) modes of the WGM resonator with the same loss rate $\kappa_o$, i.e.~the Stokes, pump and anti-Stokes mode with frequencies $\omega_{s}$, $\omega_{p}$, and $\omega_{as}$. 
When the optical free spectral range (FSR) matches the microwave frequency $\Omega_e$, resonant three-wave mixing between the microwave and adjacent optical modes arises via the cavity enhanced electro-optic interaction, with the interaction Hamiltonian
\begin{equation}
	\hat{H}_{\mathrm{eo}}/\hbar = g_{0} \hat{a}^\dagger_p \hat{a}_s \hat{b} 
	+  g_{0} \hat{a}^\dagger_p \hat{a}_{as} \hat{b}^\dagger  + h.c. ,
	\label{eq:H_I}
\end{equation}
where $\hat{a}_s$, $\hat{a}_p$, $\hat{a}_{as}$ and $\hat{b}$ are the annihilation operators for the Stokes, pump and anti-Stokes optical and microwave modes, and $g_0$ is the vacuum electro-optical coupling rate. 
A on-resonance optical pump enhances the electro-optic interaction given by $g =\sqrt{ \bar{n} }_p g_0$, where $\bar{n}_p$ is the mean intra-cavity photon number of the pump mode.
This includes the two-mode-squeezing (TMS) interaction between the Stokes and microwave mode [cf. first term in right-hand side of Eq.~\ref{eq:H_I}] and the beam-splitter (BS) interaction between the anti-Stokes mode and microwave mode  [cf. second term in right-hand side of Eq.~\ref{eq:H_I}]. 
One figure of merit of the CEO device is the multiphoton cooperativity $C = 4 \bar{n}_p g_0^2/(\kappa_o\kappa_e)$, with $\kappa_o$ and $\kappa_e$ the loss rates of the optical and microwave modes.
The TMS or BS interaction can be chosen by selectively suppressing the counterpart via mode engineering, i.e.~by coupling the anti-Stokes or Stokes mode to an optical transverse-magnetic (TM) mode of different polarization at rate of $J_{as}$ or $J_{s}$~\cite{rueda_efficient_2016}. The interaction Hamiltonian is given by
\begin{equation}
	\hat{H}_{J}/\hbar = J_{s}\hat{a}_{s}^\dagger \hat{a}_{s,\mathrm{tm}} +J_{as}\hat{a}_{as}^\dagger \hat{a}_{as,\mathrm{tm}} + h.c. ,
\end{equation}
with $\hat{a}_{s,\mathrm{tm}}$ and $\hat{a}_{as,\mathrm{tm}}$  the annihilation operators for the TM modes of frequency  $\omega_{s}$ and $\omega_{as}$.

Figure~\ref{fig:1}(c) shows the optical reflection characterization of one TE mode family of our EO device around 1550~nm with similar total loss rate $\kappa_o/2\pi \approx 26\, \mathrm{MHz}$.
We note that, all modes are re-centered to the individual TE mode resonance.
The TE modes are parametrically coupled to a microwave mode with loss rate $\kappa_e/2\pi\sim10\,\mathrm{MHz}$, whose frequency is adjusted to match the FSR. 
Mode 4 is strongly coupled to a TM mode of similar frequency with rate $J/2\pi\sim26\,\mathrm{MHz}$, which manifests as a split mode for anti-Stokes or Stokes scattering suppression when pumping mode 3 or 5 respectively.
More details regarding mode characterizations are in Supplementary Information (SI), including optical losses and mode separations. 

In the following we present temporal and spectral coherent dynamical response measurements in the pulsed regime. As shown in Fig.~\ref{fig:1}(d),
a strong optical pump pulse of duration $\tau$  is sent to the EO device, together with a weak continuous probing field around the microwave or optical (Stokes or anti-Stokes) resonance, to probe the dynamical back-action during the pulse.
We introduce the normalized probing field reflection between pump pulse on and off 
\begin{equation}
 	 R_{j}(\omega) = |S_{jj}(\omega)/S_{jj,\mathrm{off}}(\omega)|^2,
 	 \label{eq:R}
\end{equation}
with the reflection scattering parameters $S_{jj}(\omega)$, i.e. the output and input field amplitude ratio for mode $\textit{j}\in(e,o)$.

In Fig.~\ref{fig:1}(e), we show a typical normalized reflection coefficient  over time with on-resonance probing in different mode configurations for a pump pulse of duration $\tau = 250\,\mathrm{ns}$ and peak power of $\sim 500\, \mathrm{mW}$.
\begin{figure*}[th]
	\includegraphics[scale=1]{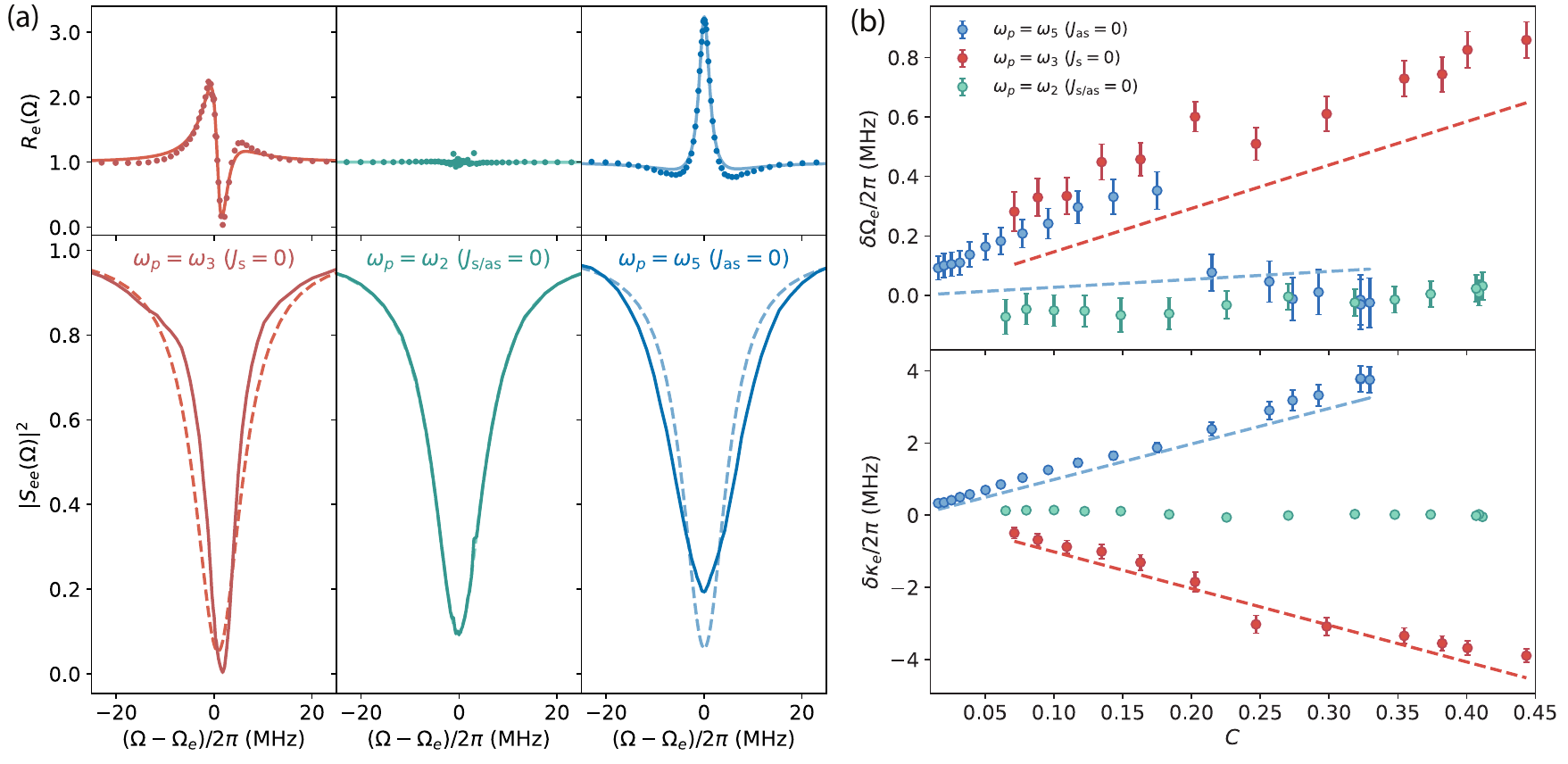}
	\caption{
		\textbf{Stationary dynamical back-action to the microwave mode.} 		
		A power sweep is conducted in each pump configuration. A joint fit of the stationary $R_e(\Omega)$ is performed with the original microwave linewidth as a shared parameter, and the linewidth and frequency change for each power as remaining fitting parameters.	
		\textbf{a}, Microwave response measurements with the same pump power as in Fig.~\ref{fig:1}(e). 
		The upper panel shows the stationary $R_e(\Omega)$  as dotted lines, with fitting curves as solid lines.
		The lower panel shows the reconstructed microwave reflection $|S_{ee}(\Omega)|^2$ with the pump on (solid curve) and off (dashed curve) using obtained parameters from the joint fit.
		\textbf{b}, Fitted microwave frequency shift and linewidth change versus cooperativity $C$.
		Dashed lines are theoretical curves incorporating the full dynamical back-action model, using fitting parameters from the corresponding coherent optical response [cf. Fig.~\ref{fig:3}], including imperfect frequency detunings. Error bars represent the 95\% confidence interval of the fit.
}
	\label{fig:2}
\end{figure*}
In the symmetric case, i.e.~mode 2 as pump mode with $J_{\mathrm{s/as}}=0$, the electro-optical dynamical back-action to the microwave mode is in principle evaded. Due to balanced Stokes and anti-Stokes scattering, the microwave susceptibility remains the same, 
\begin{equation}
	\chi_e(\Omega) = 1/(\kappa_e/2 - i \Omega).
\end{equation} 
Interestingly, the optical susceptibilities around the Stokes and anti-Stokes mode frequencies are modified, 
\begin{equation}
	\chi_{o,\mathrm{s/as}} (\Omega)	= \frac{1}{\chi_o(\Omega)^{-1}\mp g^{2}/( \chi_e(\Omega)^{-1} \pm g^2 \chi_o(\Omega) )},
\end{equation}
with $\chi_{o}(\Omega) = 1/(\kappa_{o}/2 - i \Omega)$ the  optical susceptibility.
The constructive and destructive interferences between the probing field and the electro-optical interaction result in electro-optically induced absorption (EOIA) around the Stokes mode and electro-optically induced transparency (EOIT) around the anti-Stokes mode. 
Similar dynamics has been reported previously in cavity optomechanics~\cite{safavi-naeini_electromagnetically_2011,weis_optomechanically_2010} and magnomechanics~\cite{potts_dynamical_2021}, which however only arises in the presence of dynamical back-action~\cite{aspelmeyer_cavity_2014}.
As shown in Fig.~\ref{fig:1}(e) (upper left), the microwave on-resonance reflection responds instantaneously to the arriving pump pulse,
and continues to drift even after the pulse is off ($t>250\,\mathrm{ns}$). 
Such excess back-action is negligible,
with less than $3\%$ deviation in $R_e(\Omega_e)$.
In Fig.~\ref{fig:1}(e) (lower left), the optical on-resonance Stokes (anti-Stokes) reflection decreases (increases) when the optical pulse arrives and restores instantaneously after the pulse is off.

In addition, we consider the Stokes case with mode 3 as pump mode ($J_s=0$), and the anti-Stokes case with mode 5 as pump mode ($J_{as}=0$).
Coherent electro-optical DBA results in a modified microwave susceptibility,
\begin{equation}
	\chi_{e,\mathrm{s/as}} (\Omega)	=\frac{1} {\chi_e(\Omega)^{-1} \mp g^{2} \chi_o(\Omega) }.
	\label{eq:chie_asy}
\end{equation}
DBA on the Stokes (Stokes case) or the anti-Stokes (anti-Stokes case) mode  results in the modified susceptibility,
\begin{equation}
	\chi_{o,\mathrm{s/as}} (\Omega)	= \frac{1}{\chi_o(\Omega)^{-1}\mp g^{2} \chi_e(\Omega)},
	\label{eq:chio_asy}
\end{equation}
assuming $4 J^2_{as/s} \gg \kappa_o \kappa_{o,\mathrm{tm}}$, with $\kappa_{o,\mathrm{tm}}$ the TM mode loss rate.
In both cases, Eq.~\ref{eq:chie_asy} and Eq.~\ref{eq:chio_asy} are 
symmetric under interchange of microwave and the optical probing mode, which enables mutual probing of the optical and microwave field with its counterpart.
In the normal dissipation regime, i.e.~$\kappa_o \gg\kappa_e$,  the microwave mode undergoes effective narrowing (broadening) in the Stokes (anti-Stokes) case, while the Stokes (anti-Stokes) probing field undergoes EOIA (EOIT), due to the constructive (destructive) interference between the probe field and the electro-optical interaction.
In the reversed dissipation regime, i.e.~$\kappa_o \ll\kappa_e$, the microwave mode experiences EOIA (or EOIT), while the optical Stokes (anti-Stokes) mode linewidth is effectively narrowed (broadened).
\begin{figure*}[th]
	\includegraphics[scale=1]{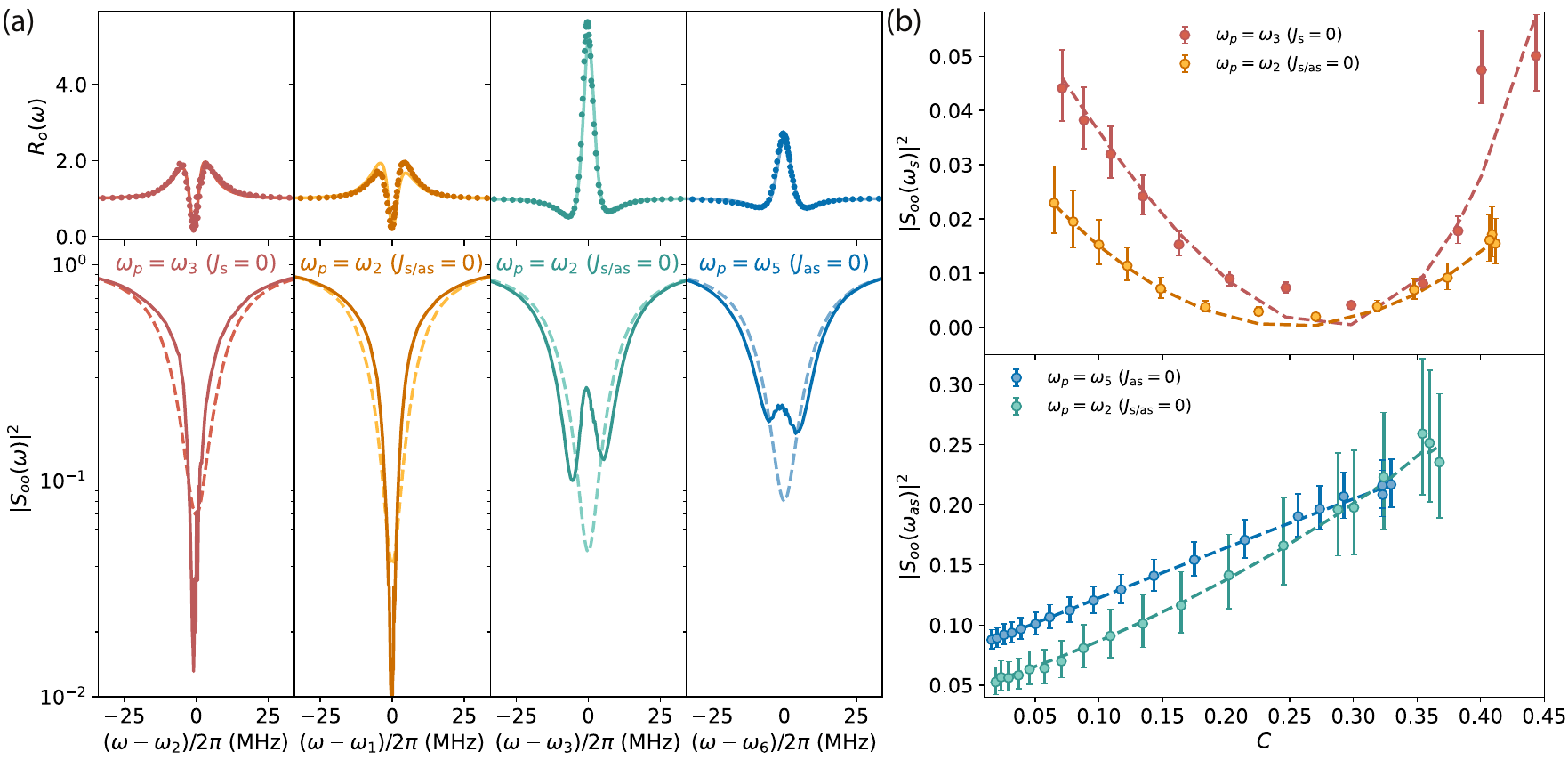}
	\caption{
		\textbf{Stationary electro-optically induced absorption and transparency}, with Stokes mode probing in the Stokes and symmetric cases, while anti-Stokes mode probing in the symmetric and anti-Stokes cases. 
		In each probing configuration, a pump power sweep is conducted, and a joint fit of stationary $R_o(\omega)$ to the full dynamical back-action model is performed.
		\textbf{a}, Measurements with same pump power as in Fig.~\ref{fig:1}(e), with the two left panels for Stokes mode probing and the two right panels for anti-Stokes mode probing. 
		The upper panel shows $R_o(\omega)$ (dotted lines) with fitting curves (solid lines). 
		The lower panel shows reconstructed optical reflection $|S_{oo}(\omega)|^2$ with pulse on (solid curve) and off (dashed curve) in logarithmic scale, which demonstrates EOIA in the Stokes case and EOIT in the anti-Stokes case. 
		\textbf{b}, The upper panel shows $|S_{oo}(\omega_s)|^2$ for the two Stokes mode probing cases, while 
		the lower panel shows $|S_{oo}(\omega_{as})|^2$ for the two anti-Stokes mode probing cases.
		The corresponding theoretical curves 
		are shown as dashed lines. Error bars indicate two standard deviations.
	}
	\label{fig:3}
\end{figure*}
The temporal on-resonance dynamics in the Stokes and anti-Stokes cases are shown in the right panel of Fig.~\ref{fig:1}(e). 
Similar to the symmetric case, the Stokes mode undergoes EOIA in the Stokes case, while the anti-Stokes mode undergoes EOIT in the anti-Stokes case.

\noindent
\textbf{Stationary Dynamical Back-action}
As shown in Fig.~\ref{fig:1}(e), the on-resonance normalized reflections remain stationary before and in the middle ($t\sim200\, \mathrm{ns}$) of the pulse.
We reconstruct the coherent stationary spectral response by sweeping the probe tone frequency around the probing mode resonance, and perform a pump pulse power sweep in each configuration.
 
To construct the microwave response, we perform a joint fit of the stationary $R_e(\Omega)$  for different powers, and obtain the individual microwave linewidth and frequency change.
The upper panel of Fig.~\ref{fig:2}(a) shows the stationary spectral response $R_e(\Omega)$ in three different pump configurations, with the same pump pulse power as in Fig.~\ref{fig:1}(e).
$R_e(\Omega)$ remains unchanged due to the balanced Stokes and anti-Stokes scattering in the symmetric case (center), while it changes dramatically around the mode resonance due to strong dynamical back-action in the two asymmetric cases.
The lower panel of Fig.~\ref{fig:2}(a) shows the measured microwave reflection scattering parameter $|S_{ee}(\Omega)|^2$ with pulse on (off)  as solid (dashed) lines, indicating microwave 
linewidth narrowing and a slight frequency increase in the Stokes case ($\omega_p = \omega_3$) and linewidth broadening in the anti-Stokes case ($\omega_p = \omega_5$) with an increased on-resonance reflection.
\begin{figure*}[th]
	\includegraphics[scale=1]{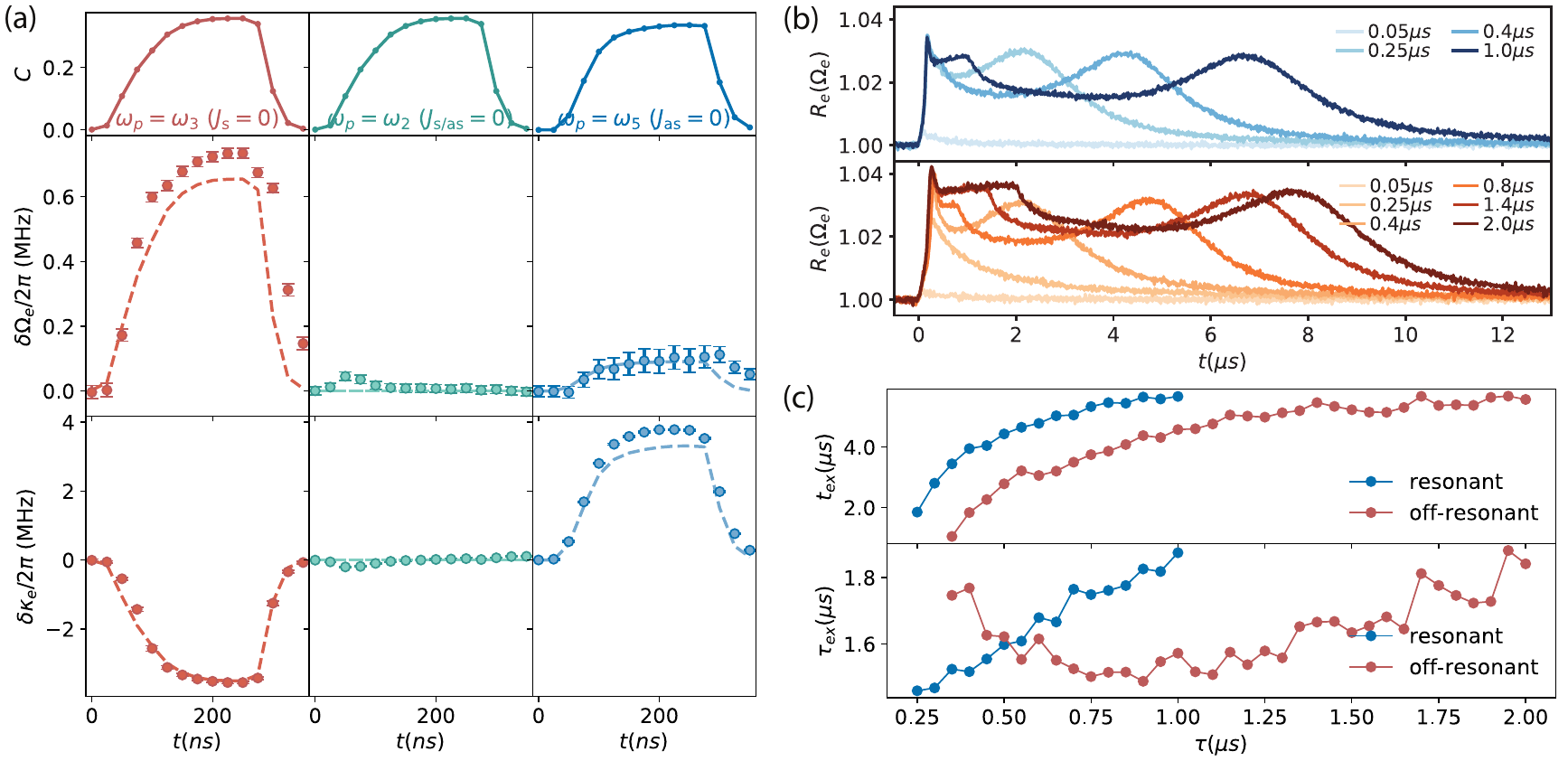}
	\caption{
		\textbf{Transient dynamical back-action to the microwave mode.} 
		\textbf{a}, Instantaneous coherent microwave responses during the pump pulse obtained for the same pump power as in Fig.~\ref{fig:1}(e).
		In each pump configuration, we perform a joint fit of $R_o(\omega)$ over the time of the pulse with the full dynamical back-action model, and extract the time dependent $C(t)$ (upper panel).
		Similarly, we perform a joint fit of $R_e(\Omega)$ to obtain the frequency and linewidth changes (dotted lines) over time, with corresponding theoretical curves (dashed lines) using the above fitted parameters from the optical coherent response.
		Error bars represent the 95\% confidence interval of the fit.
		\textbf{b}, Delayed excess back-action to the microwave mode in the resonant condition ($\Omega_e= \mathrm{FSR}$, upper panel) and the off-resonant condition ($\Omega_e\neq \mathrm{FSR}$, lower panel), in the symmetric mode configuration.
		Different curves correspond to different pulse lengths for a constant pump power similar to Fig.~\ref{fig:1}(e).
		Delayed excess back-action still exists after the pulse, and results in a bounce in $R_e(\Omega_e)$ after  $t_{\mathrm{ex}}$ of several $\mu s$, which then  decreases exponentially to 1 with the time constant $\tau_{\mathrm{ex}}$. 
		\textbf{c}, Extracted bounce time $t_{\mathrm{ex}}$ after the pulse is off and the mean decay time  $\tau_{\mathrm{ex}}$ as a function of pulse length. The blue and red curves correspond to the resonant and off-resonant cases.
	}
	\label{fig:4}
\end{figure*}

In Fig.~\ref{fig:2}(b), we show the extracted microwave frequency ($\delta\Omega_e$) and linewidth ($\delta\kappa_e$) change in the power sweep, for each pump configuration.
The corresponding microwave response fitting curves are shown in Fig.~\labelcref{figSI:1,figSI:2,figSI:3,figSI:4} in the SI.
In the symmetric case  ($\omega_p = \omega_2$), no evident frequency or linewidth change is observed due to the evaded back-action.
In the anti-Stokes case ($\omega_p = \omega_5$) the microwave linewidth increases linearly with $C$, while it decreases in the Stokes case  ($\omega_p = \omega_3$).
The theoretical curves for both asymmetric cases match very well with experimental results,
using a full dynamical back-action model incorporating optical response fitting parameters including imperfect frequency detunings [cf.~Fig.~\ref{fig:3}(b)].
In the anti-Stokes case, we observe a minuscule deviation in the microwave frequency shift of $\sim10^{-4}\Omega_e$. This can be explained by the small detuning uncertainties (sub-MHz) as discussed in the SI~\ref{sec:SIMethodTheory}, probably due to photorefractive~\cite{xu_photorefractioninduced_2021,xu_mitigating_2021}  or quasi-particles effects~\cite{mirhosseini_superconducting_2020,xu_lightinduced_2022}.

As shown in the upper panel of Fig.~\ref{fig:3}(a), we perform a joint fit of the stationary $R_{o}(\omega)$  in each probing configuration, i.e.~Stokes mode probing in the Stokes and symmetric cases while anti-Stokes mode probing in the symmetric and anti-Stokes cases.
This allows us to extract $C$, $\kappa_o$, and external coupling rate $\kappa_{o,\mathrm{ex}}$ in each probing configuration.
The detailed optical response fitting curves are show in Fig.~\labelcref{figSI:1,figSI:2,figSI:3,figSI:4}.
In the lower panel of Fig.~\ref{fig:3}(a),
we show the reconstructed optical reflection efficiency $|S_{oo}(\omega)|^2$ with pulse on and off as solid and dashed lines.
The Stokes mode probing (left two panels) reveals similar EOIA for the Stokes and symmetric cases when the pump pulse is on, while the anti-Stokes mode probing (right two panels) indicates similar EOIT for the symmetric and anti-Stokes cases.
In Fig.~\ref{fig:3}(b), we show the on-resonance reflection efficiency versus $C$ in different probing configurations with theoretical curves shown as dotted lines.
In the upper panel, $|S_{oo}(\omega_s)|^2$ at the Stokes mode resonance first approaches zero and then increases with $C$ due to EOIA.
In the lower panel, $|S_{oo}(\omega_{as})|^2$ at the anti-Stokes resonance increases slowly as $C$ increases due to EOIT. 
We note that, the different on-resonance $|S_{oo}|^2$ at low $C$ is due to the slightly different external coupling efficiency of the optical modes. 
To capture the stationary electro-optical dynamics, the effective $C$ is limited to $\sim0.5$ due to the Kerr nonlinearity~\cite{sahu_quantumenabled_2022}, which depends on the power and duration of the applied pulse and results in optical parametric oscillation in the optical resonator~\cite{kippenberg_kerrnonlinearity_2004}.
With further improvement of $\kappa_{e}$ and $g_0$, the device can enable parametric amplification of the microwave and optical Stokes signal for $C\gg1$.

\noindent
\textbf{Transient Dynamical Back-action}
Emerging quantum applications of CEO devices, such as ultra-low noise microwave-optical quantum transduction and entanglement generation, require strong optical pump pulses to reach near unity $C$~\cite{xu_bidirectional_2021,sahu_quantumenabled_2022}.
A detailed understanding of the transient response of CEO devices is therefore crucial for complex measurement protocols in the quantum limit.

In Fig.~\ref{fig:4}(a), we show the transient response of the microwave mode in different pump configurations with the same power as in Fig.~\ref{fig:1}(e).
Within each pump configuration, we perform a joint fit of $R_o(\omega)$ over the pulse incorporating the full DBA model, with $C(t)$ and imperfect detunings as free parameters, as explained in SI~\ref{sec:INEXBA}.
When the optical pump pulse arrives, the fitted $C(t)$ increases smoothly in the beginning, reaches stationary value in the middle,
and slowly decreases to zero after the pulse.
In the middle and lower panel, 
we show the obtained microwave frequency and linewidth change over the pulse as dotted lines, with theoretical curves as dashed lines. 
The small blue shift of the microwave mode in the two asymmetric mode configurations is due to imperfect detunings (sub-MHz) as explained in SI~\ref{sec:SIMethodTheory}.
The linewidth change follows closely the predicted coherent electro-optical dynamical back-action, i.e.~narrowing in the Stokes case while broadening in the anti-Stokes case.
In the symmetric case, a very slight excess frequency drift ($\sim10^{-5}\Omega_e$) and linewidth change ($\sim10^{-2}\kappa_e$)
indicate a finite amount of instantaneous excess back-action to the microwave mode in the beginning and at the end of the pulse, due to the loading and unloading of the optical pump field.
We note that, similar instantaneous excess back-action also appears in the optical coherent response, which results in an estimated detuning jiggle (sub-MHz) during the pulse, as shown in Fig.~\ref{figSI:InsBA} in SI~\ref{sec:INEXBA}.

\noindent
\textbf{Delayed Excess Back-action}
After the pump pulse, a small amount of excess back-action remains in the microwave mode for a few $\mu s$, while it ceases immediately in the optical mode.
The details of the extracted microwave frequency and linewidth drift at different $C$ over time in the symmetric case [cf. Fig.~\ref{fig:4}(a)] are discussed in Fig.~\ref{figSI:DelayBA} in SI~\ref{sec:ExDelayedBA}.
In Fig.~\ref{fig:4}(b), we show instead a comparison between two different resonant conditions, i.e.~the resonant case ($\Omega_e=\mathrm{FSR}$) and the off-resonant case ($\Omega_e\neq\mathrm{FSR}$, by detuning the microwave resonance frequency), in the symmetric mode configuration ($\omega_p = \omega_2$).
In the upper panel, we show $R_e(\Omega_e)$ over time using a similar pulse power as in Fig.~\ref{fig:1}(e), for different pulse lengths $\tau$.
The off-resonant case shown in the lower panel ($\Omega_e- \mathrm{FSR} = 2\pi\times100\,\mathrm{MHz}$) results in a similar microwave response,
which rules out the electro-optical interaction as the main origin of the induced perturbation.
For short pump pulses (e.g. below 200\,ns), $R_e(\Omega_e)$ decreases right after the pulse and restores slowly to unity in both cases.
For longer pulses (e.g. above 300\,ns), 
$R_e(\Omega_e)$ reveals an unanticipated delayed back-action, 
which decreases in the beginning and exhibits a bounce around $t_{\mathrm{ex}}$ (several $\mu s$) after the pulse.
As the pulse length increases, $t_{\mathrm{ex}}$ increases accordingly, which indicates an 
integrated optical pulse energy dependent 
excess mechanism that changes the microwave response - predominantly the mode frequency as explained in SI~\ref{sec:SIMethodTheory} and~\ref{sec:ExDelayedBA} - as corroborated by a pump power sweep with similar results [cf. Fig.~\ref{figSI:DelayBA}]. 
After the bounce, $R_e(\Omega_e)$ continues to decrease exponentially to unity with a time constant $\tau_{\mathrm{ex}}$.
In Fig.~\ref{fig:4}(c), we show the extracted $t_{\mathrm{ex}}$ and $\tau_{\mathrm{ex}}$ from the fitted time dependence for different pump pulse lengths as shown in Fig.~\ref{figSI:5}.
In cases, $t_{\mathrm{ex}}$ increases versus pulse length $\tau$ and saturates at $\sim 6\, \mu s$ for long pulse lengths above $\sim1\,\mu s$, while the resonant excess back-action arrives later than the off-resonant one most likely related to electro-optical interaction.
While the fitted decay time $\tau_{\mathrm{ex}}$ exhibits slightly different relations to the pulse length in both cases, the fitted value is quite similar ($\sim 1.6 \mu s$), indicating a general underlying mechanism, which requires further investigation, e.g. light induced quasi-particles~\cite{mirhosseini_superconducting_2020,xu_lightinduced_2022} or photo-refractive effects~\cite{xu_photorefractioninduced_2021,xu_mitigating_2021}.

It is important to point out that the observed frequency shifts and linewidth changes are only on the order of $100 \mathrm{kHz}$, i.e.~ $10^{-5}$ of the microwave resonant frequency. Moreover, our CEO device revives completely only tens of $\mu s$ after the pulse.
Nevertheless, in all the presented experiments we adopt a repetition time of 10\,ms for all pump configurations (500 ms in the Stokes case) to avoid optical heating.
Both stationary and transient coherent dynamics are independent from the different repetition rates.
We note that, the low repetition rate is important to remain in the quantum back-action dominated regime for microwave-optics entanglement generation~\cite{sahu_2022}.

\vspace{0.25cm}
\noindent
\textbf{Discussion}\\
We have demonstrated coherent optical control of a microwave cavity in the pulsed regime in a multimode EO device at millikelvin temperature with near-unitary cooperativity.
Both the stationary and instantaneous response of the microwave and optical probing field agree very well with coherent DBA theory, except for a very small and for many applications negligible excess back-action that is in fact surprisingly small given the large optical photon energy compared to the small superconducting gap of aluminum.

The presented coherent optical control of a superconducting microwave cavity mode confirms the compatibility between optical light and superconducting microwave circuits in our device, and enters a new strong interaction regime of quantum electro-optics. It also paves the way for a wide range of quantum applications beyond microwave-optical transduction~\cite{han_microwaveoptical_2021,sahu_quantumenabled_2022}, ranging from entanglement generation~\cite{sahu_2022}, optically driven masing and squeezing, to quantum thermometry~\cite{weinstein_observation_2014a} and precision measurement of microwave radiation beyond the standard quantum limit~\cite{shomroni_optical_2019}.
Our EO device offers great compatibility to cryogenic microwave circuits, and represents a promising platform for the generation of nonclassical microwave-optical correlations to realize a distributed quantum network between superconducting quantum processors~\cite{wehner_quantum_2018,arute_quantum_2019}.

\vspace{0.25cm}
\noindent
\normalsize
\textbf{Methods}\\
\noindent
\textbf{Device Characterization}\\
The multimode CEO device we use in our experiments consists of a millimeter-sized $\mathrm{LiNbO_3}$ whispering gallery mode resonator (WGMR) and a superconducting aluminum cavity at mK stage in a Bluefors dilution refrigerator, as reported previously in Refs.~\cite{hease_bidirectional_2020,sahu_quantumenabled_2022}.
The optical modes with optimal TE mode coupling are characterized individually using laser piezo scanning, whose normalized reflection are shown in Fig.~\ref{fig:1}(c) with Lorentzian fit, e.g. for mode 1, 2, and 3.
The optical resonances are $\sim1550\mathrm{nm}$, with similar fitted total linewidth of $\kappa_o/2\pi\sim 26\mathrm{MHz}$, of which the external coupling rate is $\kappa_{o,\mathrm{ex}}/2\pi\sim 10\mathrm{MHz}$.
We note that, light couples to the WGMR via a diamond prism, with imperfect spatial field mode overlap $\Lambda=0.83$. 
For simplicity, throughout our work, we include the effective $\Lambda$ factor in $\kappa_{o,\mathrm{ex}}$.
For mode 4, 5 and 6, coupling between the TE and TM mode results in mode splitting or distortion.
The mode with largest splitting, i.e.~mode 4, is adopted as the split mode in the asymmetric case.
Due to the slight splitting in mode 5, effective $C$ is slightly reduced, as evident in Fig.~\ref{fig:4}(a) with same pump pulse power.
Because of the large frequency difference and relative weak coupling between the TE and TM mode in mode 6, we approximate it as a single TE mode in the main text.
Depending on the specific pump configuration, the microwave cavity frequency is adjusted to match the optical pump and probe mode separation. 
The microwave mode has similar total loss rate $\kappa_e/2\pi\sim10\mathrm{MHz}$ with external coupling rate $\kappa_{e,\mathrm{ex}}/2\pi \sim 4\mathrm{MHz}$.

\noindent
\textbf{Data Analysis}\\
The spectral normalized reflection for the probing field is given by,
\begin{equation}
	R_j(\omega) = |S_{jj}(\omega)/S_{jj,\mathrm{off}}(\omega)|^2,
\end{equation}
where $S_{jj}(\omega)$ and $S_{jj,\mathrm{off}}(\omega)$ are the reflection coefficient ($S_{11}$ parameter) of the probing field \textit{j} with pulse on and off.
In the absence of the pump pulse, the output photon number of the probing field takes the form
\begin{equation}
	\bar{n}_{\mathrm{out,off}}(\omega) = \bar{n}_{\mathrm{in}}(\omega)  |S_{jj,\mathrm{off}}(\omega)|^2 \eta_d(\omega),
\end{equation}
where $\bar{n}_{\mathrm{in}}(\omega)$ and $\eta_d(\omega)$ are the frequency dependent input photon number and the detection efficiency.
After the pump pulse arrives, the output photon number of the probing field is modified to,
\begin{equation}
	\bar{n}_{\mathrm{out}}(\omega) = \bar{n}_{\mathrm{in}}(\omega)  |S_{jj}(\omega)|^2 \eta_d(\omega).
\end{equation}
For long repetition time as in our experiments, the coherent response of the probing field restores to the state before the pulse starts, where
we approximate $S_{jj,\mathrm{off}}(\omega)$ to $S_{jj}(\omega)|_{t=0}$.

In the experiments, the weak coherent RF signal from the down-converted microwave or optical field $I_j(t)$ is located at 40MHz, more than 10 dB above the noise floor, due to the low noise amplification using HEMT amplifier or optical balanced heterodyne detection.
We perform digital down-conversion of the time-domain data at 40MHz for each probing frequency, where the averaged voltages over the pulses are adopted to obtain the mean power.
We can track the normalized reflection coefficient over time by scanning the probe field frequency,
\begin{equation}
	R_j(\Omega+\Omega_{\mathrm{LO},j}) = \frac{ \bar{P}_{\mathrm{out},j}(\Omega)}{ \bar{P}_{\mathrm{out},j}(\Omega)|_{t=0}},
\end{equation}
with $\Omega_{\mathrm{LO},j}$ the LO frequency and $\bar{P}_{\mathrm{out},j}$ the averaged power of the RF field from digital down-conversion.
Typical obtained on-resonance $R_j(\omega)$ in time domain are shown in Fig.~\ref{fig:1}(e).
In this way, we avoid the complicated system calibration and frequency dependence on the input and detection sides.

\vspace{0.25cm}
\noindent
\normalsize
\textbf{Data Availability}\\
The code and raw data used to produce the plots in this paper are available at a Zenodo open-access repository under the link https://doi.org/10.5281/zenodo.7936405.\\

%

\vspace{0.25cm}
\noindent
\normalsize
\textbf{Acknowledgments}\\
This work was supported by the European Research Council under grant agreement no.~758053 (ERC StG QUNNECT) and the European Union's Horizon 2020 research and innovation program under grant agreement no.~899354 (FETopen SuperQuLAN). L.Q.~acknowledges generous support from the ISTFELLOW programme. W.H.~is the recipient of an ISTplus postdoctoral fellowship with funding from the European Union's Horizon 2020 research and innovation program under the Marie Sk\l{}odowska-Curie grant agreement no.~754411. G.A.~is the recipient of a DOC fellowship of the Austrian Academy of Sciences at IST Austria.\\

\vspace{0.25cm}
\noindent
\normalsize
\textbf{Author Contributions}\\
L.Q. conceived the idea for the experiment.
L.Q., and R.S. performed the experiments together with W.H. and G.A. L.Q. developed the theory and performed the data analysis. 
The manuscript was written by L.Q. with assistance from all authors. J.M.F. supervised the project.\\

\vspace{0.25cm}
\noindent
\normalsize
\textbf{Competing Interests} \\
The authors declare no competing interests.


\makeatletter
\close@column@grid
\clearpage

\onecolumngrid
\begin{center}
	\textbf{\large 
		\@title\\[0.5cm]
		Supplementary Information\\[.5cm]
	}
	Liu Qiu,$^{1,*\dagger}$ Rishabh Sahu,$^{1,*}$ William Hease,$^1$ Georg Arnold,$^1$ and Johannes M. Fink$^{1,\ddagger}$\\[.1cm]
	{\itshape 
		$^1$Institute of Science and Technology Austria, am Campus 1, 3400 Klosterneuburg, Austria
		}\\
	$^\dagger$electronic address: liu.qiu@ist.ac.at\\
	$^\ddagger$electronic address: johannes.fink@ist.ac.at\\
	(Dated: \@date)\\[1cm]
\end{center}

\title{Dynamical back-action in multimode cavity electro-optics}
\author{Liu Qiu}
\email{liu.qiu@ist.ac.at}
\thanks{These two authors contributed equally}
\author{Rishabh Sahu}
\thanks{These two authors contributed equally}
\author{William Hease}
\author{Georg Arnold}
\author{Johannes M. Fink}
\email{johannes.fink@ist.ac.at}

\makeatother

\setcounter{equation}{0}
\setcounter{figure}{0}
\setcounter{table}{0}
\setcounter{page}{1}
\renewcommand{\theequation}{S\arabic{equation}}
\renewcommand{\thefigure}{S\arabic{figure}}
\renewcommand{\thetable}{S\arabic{table}}
\renewcommand{\bibnumfmt}[1]{[S#1]}
\renewcommand{\citenumfont}[1]{S#1}


\subsection{Theoretical model}\label{sec:SIMethodTheory}
The total interaction Hamiltonian of the multimode cavity electro-optical system is given by $\hat{H}_{I} = \hat{H}_{\mathrm{eo}} +\hat{H}_J $, where
$\hat{H}_{\mathrm{eo}}/\hbar = g_{0} \hat{a}^\dagger_p \hat{a}_s \hat{b} 
+  g_{0} \hat{a}^\dagger_p \hat{a}_{as} \hat{b}^\dagger   + h.c. $,
and 
$	\hat{H}_{J}/\hbar = J_{s}\hat{a}_{s}^\dagger \hat{a}_{s,\mathrm{tm}} +J_{as}\hat{a}_{as}^\dagger \hat{a}_{as,\mathrm{tm}} + h.c.$.
Here, $\hat{a}_s$, $\hat{a}_p$, $\hat{a}_{as}$ and $\hat{b}$ are the annihilation operators for the Stokes, pump and anti-Stokes optical modes and microwave mode, and $g_0$ is the vacuum electro-optical coupling rate. 
$\hat{a}_{s,\mathrm{tm}}$ and $\hat{a}_{as,\mathrm{tm}}$ are the annihilation operators for the corresponding TM mode of the Stokes and anti-Stokes mode.

We define the susceptibility of microwave or optical mode as,
\begin{equation}
	\chi_j(\Omega) = \frac{1}{\kappa_j/2 - i \Omega},
\end{equation}
where \textit{j} is e or o for microwave or optical mode.

The strong optical \textit{on-resonance} pump results in photon enhanced electro-optical coupling rate $g = \sqrt{\bar{n}_p} g_0 $, with $\bar{n}_p$ the mean photon number of the pump mode.
Scattered Stokes and anti-Stokes sidebands are located on the lower and upper side of the pump by $\Omega_e$.
In practice,  the Stokes and anti-Stokes sidebands are detuned from the Stokes and anti-Stokes mode by $\delta_{\mathrm{s}}$ and $\delta_{\mathrm{as}}$, mostly due to FSR and $\Omega_e$ mismatch.
In the case of on-resonance pumping and zero dispersion, we have $\delta_{\mathrm{s}} = - \delta_{\mathrm{as}}$.
For simplicity, we assume the TM mode is of the same frequency of the corresponding Stokes  or anti-Stokes mode.

We define the noise operators in the vector form, with the mode operator and input noise operator vectors,
\begin{equation}
	\begin{aligned}
		D &= \left( \hat{a}_s, \hat{a}^{\dagger}_s, \hat{a}_{as}, \hat{a}^{\dagger}_{as}, \hat{a}_{s,\mathrm{tm}}, \hat{a}^{\dagger}_{s,\mathrm{tm}},\hat{a}_{as,\mathrm{tm}}, \hat{a}^{\dagger}_{as, \mathrm{tm}}, \hat{b}, \hat{b}^{\dagger} \right)^{T}, \\
		D_{\mathrm{in}} &= \left(
		\hat{a}_{s,\mathrm{in}}, \hat{a}^{\dagger}_{s,\mathrm{in}},
		\hat{a}_{s,\mathrm{0}}, \hat{a}^{\dagger}_{s,\mathrm{0}},
		\hat{a}_{as,\mathrm{in}},\hat{a}^{\dagger}_{as,\mathrm{in}},
		\hat{a}_{as,\mathrm{0}},\hat{a}^{\dagger}_{as,\mathrm{0}},
		\hat{a}_{s,\mathrm{tm,vac}},\hat{a}^{\dagger}_{s,\mathrm{tm,vac}},
		\hat{a}_{as,\mathrm{tm,vac}},\hat{a}^{\dagger}_{as,\mathrm{tm,vac}},
		\hat{b}_{\mathrm{in}}, \hat{b}_{\mathrm{in}}^{\dagger}, 
		\hat{b}_{\mathrm{0}}, \hat{b}_{\mathrm{0}}^{\dagger} 
		\right)^{T}, 
	\end{aligned}
\end{equation}
with $\hat{a}_{j,\mathrm{in}} $ ($\hat{a}_{j,0}$) and $\hat{b}_{\mathrm{in}}$ ( $\hat{b}_{0}$) the input (intrinsic) noise operator for optical and microwave modes, and $\hat{a}_{j,\mathrm{tm,vac}}$ the TM vacuum noise.
In the rotating frame of the microwave resonance and Stokes and anti-Stokes sideband,  we obtain the full dynamics of the intracavity mode field from quantum Langevin equations in the Fourier space,
\begin{equation}
	\hat{D} = \textbf{M}\cdot  \textbf{L}\cdot \hat{D}_{\mathrm{in}},
	\label{eq:QLE}
\end{equation}
where
\begin{equation}
\textbf{M}= 
\begin{pmatrix}
	\chi^{-1}_o(\delta _s+\Omega ) & 0 & 0 & 0 & -i J_s & 0 & 0 & 0 & 0 & i g \\
0 & \!\!\!\!\!\!\!\!\chi^{-1}_o(\Omega -\delta _s) & 0 & 0 & 0 & i J_s & 0 & 0 & -i g & 0 \\
0 & 0 & \!\!\!\!\!\!\!\!\chi^{-1}_o(\delta _{\text{as}}+\Omega ) & 0 & 0 & 0 & -i J_{\text{as}} & 0 & i g & 0 \\
0 & 0 & 0 & \!\!\!\!\!\!\!\!\chi^{-1}_o(\Omega -\delta _{\text{as}}) & 0 & 0 & 0 & i J_{\text{as}} & 0 & -i g \\
-i J_s & 0 & 0 & 0 & \!\!\!\!\!\!\!\!\chi^{-1}_{o,\mathrm{tm}}(\delta _s+\Omega) & 0 & 0 & 0 & 0 & 0 \\
0 & i J_s & 0 & 0 & 0 & \!\!\!\!\!\!\!\!\chi^{-1}_{o,\mathrm{tm}}(\Omega -\delta _s) & 0 & 0 & 0 & 0 \\
0 & 0 & -i J_{\text{as}} & 0 & 0 & 0 & \!\!\!\!\!\!\!\!\chi^{-1}_{o,\mathrm{tm}}(\delta _{\text{as}}+\Omega) & 0 & 0 & 0 \\
0 & 0 & 0 & i J_{\text{as}} & 0 & 0 & 0 & \!\!\!\!\!\!\!\!\chi^{-1}_{o,\mathrm{tm}}(\Omega -\delta _{\text{as}}) & 0 & 0 \\
0 & i g & i g & 0 & 0 & 0 & 0 & 0 & \!\!\!\!\!\chi^{-1}_e(\Omega ) & 0 \\
-i g & 0 & 0 & -i g & 0 & 0 & 0 & 0 & 0 & \!\!\!\!\!\!\chi^{-1}_e(\Omega )
\end{pmatrix}^{-1}
\end{equation}
\begin{equation}
\textbf{L} = 
\left(
\begin{array}{cccccccccccccccc}
	\sqrt{\kappa _{o,\text{ex}}} & 0 & \!\!\!\! \sqrt{\kappa _{o,0}} & 0 & 0 & 0 & 0 & 0 & 0 & 0 & 0 & 0 & 0 & 0 & 0 & 0 \\
	0 &  \!\!\!\! \sqrt{\kappa _{o,\text{ex}}} & 0 &  \!\!\!\!\sqrt{\kappa _{o,0}} & 0 & 0 & 0 & 0 & 0 & 0 & 0 & 0 & 0 & 0 & 0 & 0 \\
	0 & 0 & 0 & 0 & \!\!\!\! \sqrt{\kappa _{o,\text{ex}}} & 0 &  \!\!\!\!\sqrt{\kappa _{o,0}} & 0 & 0 & 0 & 0 & 0 & 0 & 0 & 0 & 0 \\
	0 & 0 & 0 & 0 & 0 & \!\!\!\! \sqrt{\kappa _{o,\text{ex}}} & 0 &  \!\!\!\!\sqrt{\kappa _{o,0}} & 0 & 0 & 0 & 0 & 0 & 0 & 0 & 0 \\
	0 & 0 & 0 & 0 & 0 & 0 & 0 & 0 &  \!\!\!\!\sqrt{\kappa _{o,\mathrm{tm}}} & 0 & 0 & 0 & 0 & 0 & 0 & 0 \\
	0 & 0 & 0 & 0 & 0 & 0 & 0 & 0 & 0 & \!\!\!\! \sqrt{\kappa _{o,\mathrm{tm}}} & 0 & 0 & 0 & 0 & 0 & 0 \\
	0 & 0 & 0 & 0 & 0 & 0 & 0 & 0 & 0 & 0 & \!\!\!\! \sqrt{\kappa _{o,\mathrm{tm}}} & 0 & 0 & 0 & 0 & 0 \\
	0 & 0 & 0 & 0 & 0 & 0 & 0 & 0 & 0 & 0 & 0 & \!\!\!\! \sqrt{\kappa _{o,\mathrm{tm}}} & 0 & 0 & 0 & 0 \\
	0 & 0 & 0 & 0 & 0 & 0 & 0 & 0 & 0 & 0 & 0 & 0 & \!\!\!\! \sqrt{\kappa _{e,\text{ex}}} & 0 &  \!\!\!\!\sqrt{\kappa _{e,0}} & 0 \\
	0 & 0 & 0 & 0 & 0 & 0 & 0 & 0 & 0 & 0 & 0 & 0 & 0 & \!\!\!\! \sqrt{\kappa _{e,\text{ex}}} & 0 & \!\!\!\! \sqrt{\kappa _{e,0}} \\
\end{array}
\right).
\end{equation}
$\kappa_{j,0}$ and $\kappa_{j,\mathrm{ex}}$ correspond to the intrinsic loss and external coupling rate of mode $j$.
We can obtain the effective susceptibility of microwave and optical mode $\chi_{j,\mathrm{eff}}(\Omega)$ from Eq.~\ref{eq:QLE}.

The output probing field can be obtained via input-output theorem, $\hat{a}_{j,\mathrm{out}} = \hat{a}_{j,\mathrm{in}} - \sqrt{\kappa_{j,\mathrm{ex}}} \hat{a}_j$.
From the output field, we can obtain the incoherent output noise spectral density via Wiener–Khinchin theorem in different detection schemes.
In this work, we focus on the coherent response of the multimode CEO system, where the amplitude reflection efficiency \textit{in the lab frame} is given by,
\begin{equation}
	S_{jj}(\Omega+\omega_j) = 1- \eta_{j} \kappa_{j} \chi_{j,\mathrm{eff}}(\Omega),
\end{equation}
with $\eta_j = \kappa_{j,\mathrm{ex}}/\kappa_{j}$ the external coupling efficient for mode $j$.
We define the  spectral normalized reflection as the ratio of the reflection efficiency between pulse on and off, 
\begin{equation}
	R_{j}(\omega) = |S_{jj}(\omega)/S_{jj,\mathrm{off}}(\omega)|^2,
	\label{eq:SIR}
\end{equation}
with $S_{jj,\mathrm{off}}(\omega) =  1- \kappa_{j,\textrm{ex}} \chi_{j}(\Omega)$.
Before the pulse is switched on, the absent back-action leads to $R_{j}(\omega) = 1 $.
During the pulse, the coherent and excess back-action leads to modification of $R_{j}(\omega)$.
After the pulse, $R_{j}(\omega)$ restores to 1 if the repetition time is long enough.
We note that, in case of \textit{on-resonance microwave probing}, the normalized reflection
\begin{equation}
	R_e(\Omega_e) = \left|\frac{1 - \kappa_{e,\mathrm{ex}}/((\kappa_{e}+\delta\kappa_e)/2 + i  \delta \Omega_e)}{1 - 2 \eta_e} \right|^2,
	\label{eq:SIRonres}
\end{equation}
is more susceptible to the microwave frequency shift $\delta\Omega_e$ compared to the linewidth change $\delta\kappa_e$.

\subsubsection{Symmetric mode configuration}
\begin{figure}[th]
	\includegraphics[scale=1]{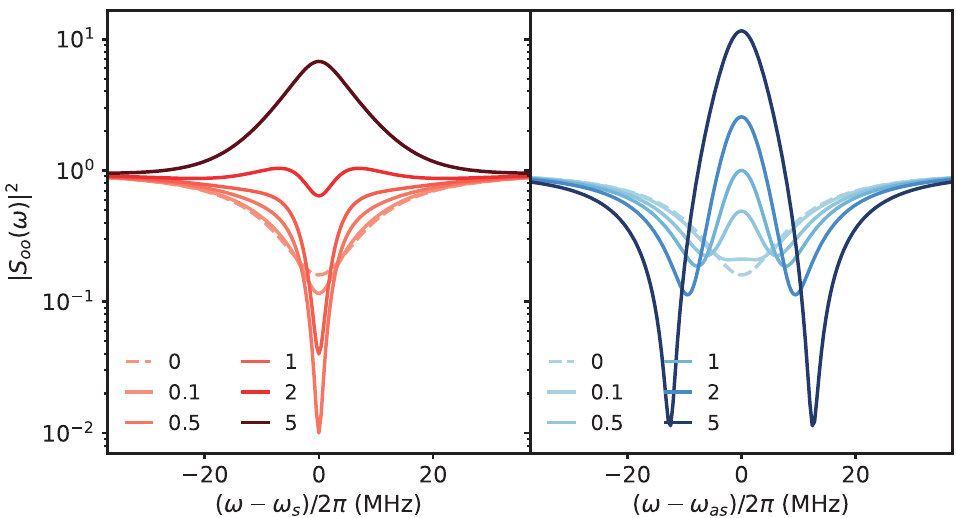}
	\caption{
		\textbf{ Theoretical coherent optical response in the symmetric case at different $C$.} The left panel corresponds to the optical probing around the Stokes mode, 
		while the right panel corresponds to optical probing around the anti-Stokes mode. (Parameters:
		$\kappa_o/2\pi=30\mathrm{MHz}$, $\eta_o=0.3$, $\kappa_e/2\pi=10\mathrm{MHz}$)
	}
	\label{figSI:Mutual_sym}
\end{figure}

For ideal detunings ($\delta_s = \delta_{\mathrm{as}} = 0$),
microwave effective susceptibility remains the same, 
\begin{equation}
	\chi_{e,\mathrm{eff}} (\Omega)	= \chi_e(\Omega),
\end{equation} 
due to evaded electro-optical dynamical back-action.
In practice, excess back-action exists where $R_e(\Omega)$  slightly deviates from 1, as shown in the main text.
Despite the absent dynamical back-action to the microwave mode, the optical susceptibility around the Stokes and anti-Stokes modes are changed, which takes the form, 
\begin{equation}
	\chi_{o,\mathrm{s/as}} (\Omega)	= \frac{1}{\chi_o(\Omega)^{-1}\mp g^{2}/( \chi_e(\Omega)^{-1} \pm g^2 \chi_o(\Omega) )}.
\end{equation}
In Fig.~\ref{figSI:Mutual_sym} , we show the theoretical curves of the optical coherent response in the symmetric case ($J_{\mathrm{s/as}}=0$) at different $C$.
For low $C$, $	\chi_{o,\mathrm{s}} (\Omega)$ and $	\chi_{o,\mathrm{as}} (\Omega)$ show similar behavior to electro-optically induced absorption (EOIA) and transparency (EOIT), due to the constructive and destructive interference between the probe field and the electro-optical interaction induced field.
As $C$ increases, both the Stokes and anti-Stokes mode probing response deviate from typical EOIA and EOIT behavior.
For example, the optical reflection coefficient can even exceed unitary around resonance for anti-Stokes mode probing.
Even in the symmetric case, the complex optical response of the multimode CEO system can be utilized for dispersion engineering of the probing field.
At large $C$ (e.g. $C \gg2$), the symmetric multimode CEO system can function as a broadband electro-optical parametric amplifier for both Stokes and anti-Stokes signals.

\subsubsection{Stokes mode configuration}
In the Stokes case, i.e.~$J_{s} = 0$, effective microwave susceptibility is given by,
\begin{equation}
	\chi_{e,s} (\Omega)	= \frac{1}
	{\chi^{-1}_e\left(\Omega\right) 
		+ g^{2}  \left(
		\frac{\chi_o(\Omega+\delta_{\mathrm{as}})}
		{1+J_{\mathrm{as}}^2 \chi_o(\Omega+\delta_{\mathrm{as}}) \chi_{o,\mathrm{tm} }(\Omega+\delta_{\mathrm{as}}) }
		-\chi_o(\Omega+\delta_s)
		\right).
	}
\end{equation}

As shown in the theoretical curves in Fig.~\ref{fig:2} and~\ref{fig:3}, dynamical back-action results in microwave frequency shift (optical-spring effect) and linewidth decrease,
\begin{equation}
	\begin{aligned}
	\delta \Omega_e &= -\frac{4 g^2 \delta _s}{\kappa _o^2+4 \delta _s^2}+ \frac{4 \delta_{as} g^2 \left(\kappa _{o, \mathrm{tm} }^2-4 J_{\text{as}}^2+4 \delta_{as}^2\right)}{8 J_{\text{as}}^2 \left(\kappa _o \kappa _{o, \mathrm{tm}}-4 \delta_{as}^2\right)+\left(4 \delta_{as}^2+\kappa _o^2\right) \left(\kappa _{o,\mathrm{tm}}^2+4 \delta_{as}^2\right)+16 J_{\text{as}}^4}\\
	\delta \kappa_e &=  -\frac{4 g^2 \kappa _o}{\kappa _o^2+4 \delta _s^2}
	+ \frac{4 g^2 \left(\kappa _{o,\mathrm{tm}} \left(\kappa _o \kappa _{o, \mathrm{tm}}+4 J_{\text{as}}^2\right)+4 \delta_{as}^2 \kappa _o\right)}{8 J_{\text{as}}^2 \left(\kappa _o \kappa _{o,\mathrm{tm}}-4 \delta_{as}^2\right)+\left(4 \delta_{as}^2+\kappa _o^2\right) \left(\kappa _{o,\mathrm{tm}}^2+4 \delta_{as}^2\right)+16 J_{\text{as}}^4},
	\end{aligned}
\label{eq:delta_s}
\end{equation}
where our experiment are in the normal dissipation regime, i.e.~$\kappa_o\gg\kappa_e$.

In the case of ideal detuning, i.e.~$\delta_{s} = \delta_{as} = 0$, 
\begin{equation}
	\begin{aligned}
		&\chi_{e,s} (\Omega)	= \frac{1}
		{\chi^{-1}_e\left(\Omega\right) 
			- g^{2}\chi_o(\Omega)  \left(1 - r_{\mathrm{as}}(\Omega)
			\right)
		}\\
		&\chi_{o,{\mathrm{s}}} (\Omega)	= \frac{1}
		{
			\chi^{-1}_o(\Omega) - g^{2}\chi_e(\Omega)/( 1 + g^2 \chi_e(\Omega) \chi_o(\Omega) r_{\mathrm{as}}(\Omega))
		}
	\end{aligned},
\end{equation}
where $r_{\mathrm{as}}(\Omega) = [1+J_{as}^2 \chi_o(\Omega) \chi_{o,\mathrm{tm}}(\Omega) ]^{-1} <1$ is the anti-Stokes and Stokes scattering rate ratio.

In the case of $4 J^2_{\mathrm{as}} \gg \kappa_o \kappa_{o,\mathrm{tm}}$, where the anti-Stokes scattering is completely suppressed, we obtain,
\begin{equation}
	\begin{aligned}
	\chi_{e,s} (\Omega)	&= \frac{1}{\chi_e(\Omega)^{-1} - g^{2} \chi_o(\Omega) }\\
	\chi_{o,\mathrm{s}} (\Omega)	&= \frac{1}{\chi_o(\Omega)^{-1} - g^{2} \chi_e(\Omega) }		
	\end{aligned},
\end{equation}
which is \textit{symmetric under interchange of microwave and the Stokes mode.}

\begin{figure}[th]
	\includegraphics[scale=1]{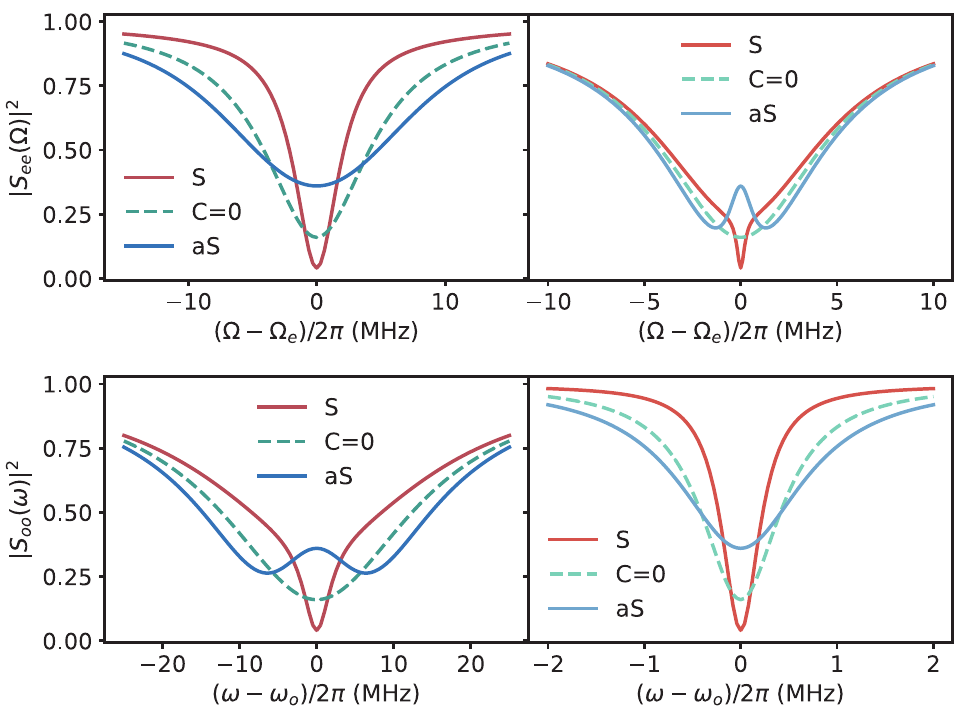}
	\caption{
		\textbf{Theoretical coherent multimode electro-optical dynamical back-action in normal and reversed dissipation regime.} 
		The left two panels correspond to the normal dissipation regime with $\kappa_o\gg\kappa_e$.
		(Parameters: $\kappa_o/2\pi=30\mathrm{MHz}$, $\eta_o=0.3$, $\kappa_e/2\pi=10\mathrm{MHz}$,  $\eta_e=0.3$, and $C=0.5$)
		The red and blue solid curves correspond to the Stokes and anti-Stokes case with $C=0.5$, while the dashed curves correspond to the original reflection coefficient with $C=0$.
		The right two panels correspond to the reversed dissipation regime with $\kappa_o\ll\kappa_e$.
		Different curves are similar to those in the left panels.
		(Parameters: $\kappa_o/2\pi=1\mathrm{MHz}$, $\eta_o=0.3$, $\kappa_e/2\pi=10\mathrm{MHz}$,  $\eta_e=0.3$, and $C=0.5$)
	}
	\label{figSI:Mutual}
\end{figure}
As seen from the red curves in Fig.~\ref{figSI:Mutual}, 
the microwave response shows 
\textit{effective narrowing in the normal dissipation regime (upper left), while EOIA in the reversed dissipation regime (upper right).
}
The optical response around the Stokes mode shows 
\textit{EOIA in the normal dissipation regime (lower left), while effective narrowing  in the reversed dissipation regime (lower right).
}
The asymmetric multimode CEO system can be adopted for "fast light" of optical (microwave) probing field in the normal (reversed) dissipation regime, with reduced group delay.

\subsubsection{anti-Stokes mode configuration}
In the anti-Stokes case, i.e.~$J_{\mathrm{as}} =0$, effective microwave susceptibility is given by,
\begin{equation}
	\chi_{e,\mathrm{as}} (\Omega)	= \frac{1}
	{\chi_e\left(\Omega\right)^{-1} 
		+ g^{2}  \left(
		\chi_o\left(\Omega+\delta_{\mathrm{as}}\right)
		-\frac{\chi_o\left(\Omega+\delta_s\right)}
		{1+J_s^2 \chi_o\left(\Omega+\delta_s\right) \chi_{o,\mathrm{tm}}\left(\Omega+\delta_s\right)}
		\right)
	}
\end{equation}
As shown in the theoretical curves in Fig.~\ref{fig:2} and~\ref{fig:3}, the dynamical back-action results in optical-spring effect and effective microwave linewidth increase, 
\begin{equation}
	\begin{aligned}
	\delta \Omega_e &= \frac{4 g^2 \delta _{\text{as}}}{4 \delta _{\text{as}}^2+\kappa _o^2}
	-\frac{4 g^2 \delta _s \left(\kappa _{o,\mathrm{tm}}^2-4 J_s^2+4 \delta _s^2\right)}{8 J_s^2 \left(\kappa _o \kappa _{o,\mathrm{tm}}-4 \delta _s^2\right)+\left(\kappa _o^2+4 \delta _s^2\right) \left(\kappa _{o,\mathrm{tm}}^2+4 \delta _s^2\right)+16 J_s^4}\\
	\delta \kappa_e &=  \frac{4 g^2 \kappa _o}{4 \delta _{\text{as}}^2+\kappa _o^2}
	-\frac{4 g^2 \left(\kappa _{o,\mathrm{tm}} \left(\kappa _o \kappa _{o,\mathrm{tm}}+4 J_s^2\right)+4 \kappa _o \delta _s^2\right)}{8 J_s^2 \left(\kappa _o \kappa _{o,\mathrm{tm}}-4 \delta _s^2\right)+\left(\kappa _o^2+4 \delta _s^2\right) \left(\kappa _{o,\mathrm{tm}}^2+4 \delta _s^2\right)+16 J_s^4},		
	\end{aligned}
\label{eq:delta_as}
\end{equation}
as our experiment is in the normal dissipation regime, i.e.~$\kappa_o\gg\kappa_e$.

In the case of ideal detuning, i.e.~$\delta_{s} = \delta_{\mathrm{as}} = 0$, 
\begin{equation}
	\begin{aligned}
		&\chi_{e,\mathrm{eff}} (\Omega)	= \frac{1}
		{\chi^{-1}_e\left(\Omega\right) 
			+ g^{2}\chi_o(\Omega)  \left(1 - r_{\mathrm{s}}(\Omega)
			\right)
		}\\
		&\chi_{o,{\mathrm{as}}} (\Omega)	= \frac{1}
		{
			\chi^{-1}_o(\Omega) +  g^{2}\chi_e(\Omega)/( 1 - g^2 \chi_e(\Omega) \chi_o(\Omega) r_{\mathrm{s}}(\Omega))
		}
	\end{aligned},
\end{equation}
where $r_{s}(\Omega) = [1+J_{s}^2 \chi_o(\Omega) \chi_{o,\mathrm{tm}}(\Omega) ]^{-1} <1$ is the Stokes and anti-Stokes scattering rate ratio.
When the anti-Stokes scattering is fully suppressed ($4 J^2_{s} \gg \kappa_o \kappa_{o,\mathrm{tm}}$), we obtain,
\begin{equation}
	\begin{aligned}
			\chi_{e,\mathrm{as}} (\Omega)	&= \frac{1}{\chi_e(\Omega)^{-1} + g^{2} \chi_o(\Omega)   }\\
			\chi_{o,\mathrm{as}} (\Omega)	&= \frac{1}{\chi_o(\Omega)^{-1} + g^{2} \chi_e(\Omega)   },
	\end{aligned}
\end{equation}
which is \textit{symmetric under interchange of microwave and the anti-Stokes mode.}

As seen from the blue curves in Fig.~\ref{figSI:Mutual}, 
the microwave response shows 
\textit{effective broadening in the normal dissipation regime  (upper left), while EOIT in the reversed dissipation regime  (upper right).
}
The optical response around the anti-Stokes mode shows 
\textit{EOIT in the normal dissipation regime (lower left), while effective broadening in the reversed dissipation regime (lower right).
}
The asymmetric multimode CEO system can be adopted for "slow light" of optical  (microwave) probing field in the normal (reversed) dissipation regime, with increased group delay.
For simplicity, we assume the same cavity coupling coefficient 0.3 for both microwave and optical modes in the theoretical calculations in Fig.~\ref{figSI:Mutual}.

\subsection{Experimental Setup}\label{sec:SISetup}
\begin{figure}[!ht]
	\includegraphics[scale=0.67]{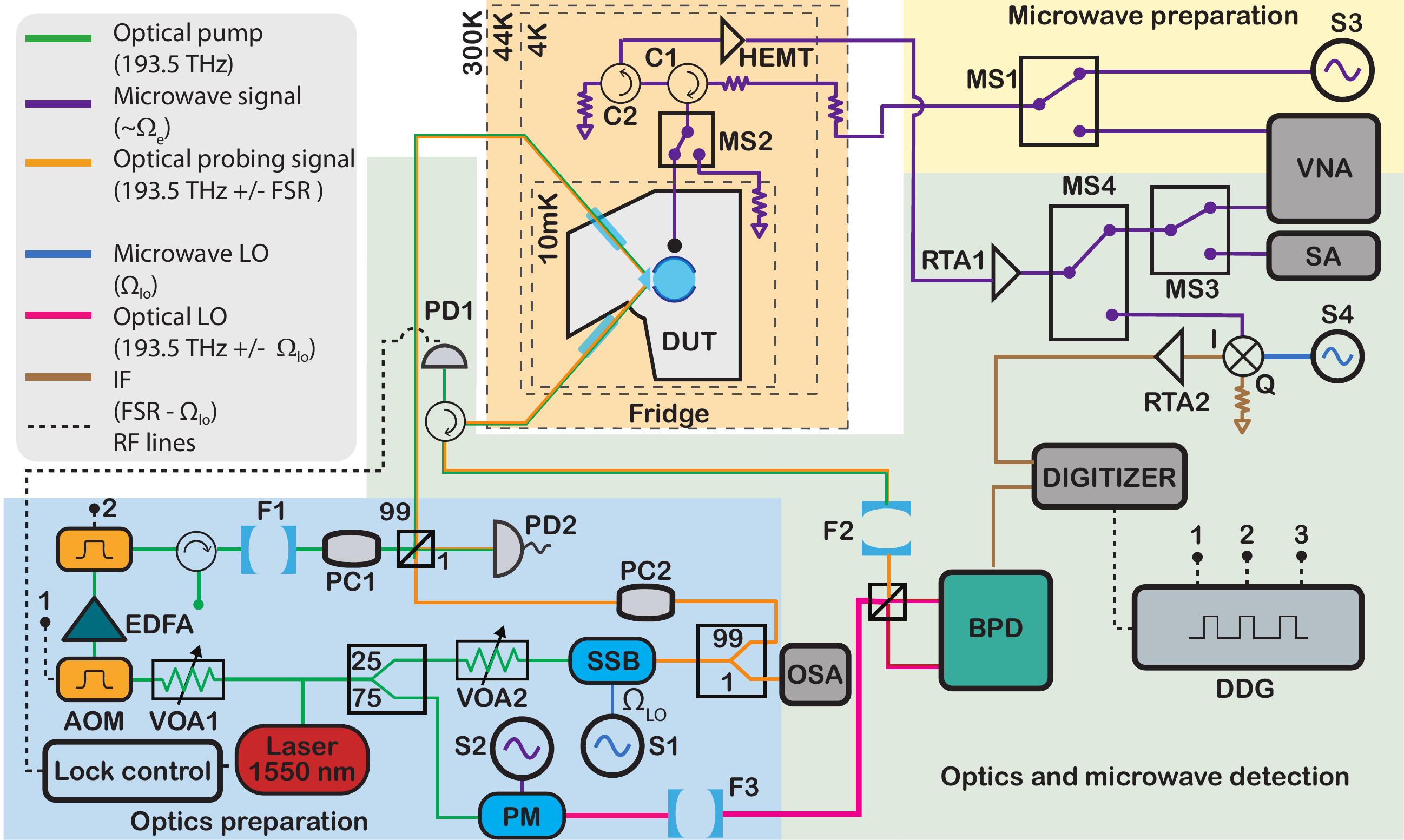}
	\caption{
		\textbf{Experimental Setup}.
		On the optical side, laser light from Toptica CTL 1550
		is divided into two equal parts - one for the optical pump, and the other for the optical signal and  local oscillator (LO).
		The optical pump (left) passes through a variable optical attenuator (VOA1), and is then sent to an acousto-optic modulator (AOM1). 
		The AOM is pulsed using a digital delay generator (DDG). 
		The generated optical pulses are amplified by an Erbium-doped fiber amplifier (EDFA).
		Output of the EDFA is sent to another AOM2, which is connected to DDG and is used to suppress the broad-band amplified spontaneous emission (ASE) noise from the EDFA. 
		The optical pump pulse is further filtered with an on-resonance filter cavity F1 (50 MHz linewidth with $\sim$15 GHz FSR), which is frequency locked to the Toptica laser. 
		The signal arm is first divided in two parts, i.e.~the optical signal and LO.
		25\% of the light in signal arm is used to produce the optical probing signal using a single sideband modulator (SSB),  continuously monitored by an optical spectrum analyzer (OSA). 
		The 75\% light on the right side of the laser is used to produce the optical LO via a phase modulator (PM). 
		The PM is operated via a microwave source S2 for efficient sideband generation and carrier suppression.
		Optical LO is sent to a filter cavity F3 to suppress the spurious optical tones from the PM. 
		The pump pulse together with the optical probing signal are sent to the dilution refrigerator (DR). 
		In the DR, light is focused via a gradient-index (GRIN) lens on the surface of a prism and coupled to the optical whispering gallery mode resonator (WGMR) via evanescent coupling. 
		Fiber polarization controllers PC1 and PC2 are adjusted for efficient TE mode coupling of the optical WGMR.
		The output light from the optical WGMR is collected by the grin lens and is sent to filter F2 (50 MHz linewidth with $\sim$40 dB suppression) to reject the strong optical pump. 
		The reflected optical pump from cavity F2 is used to lock the laser to the optical pump mode and is partially captured by photo-diode PD1 via the circulator C4. 
		The transmitted optical probing signal and optical LO are sent directly to the optical balanced heterodyne setup with a balanced photo-detector (BPD). 
		The output signal is amplified via a room temperature amplifier RTA1 before being sent to a digitizer.
		On the microwave side, the probing microwave signal is sent from microwave source S3 
		(or from the VNA for microwave mode spectroscopy) to the fridge input line via the microwave switch (MS1). 
		The input line is attenuated by 60dB with cryogenic attenuators distributed between 3 K and 10 mK to suppress room temperature microwave noise. 
		Circulator C1 redirects the reflected tone from the cavity to amplified output line, while C2 redirects noise coming in from the output line to a matched 50 $\Omega$ termination.
		Microwave switch MS2 allows to swap the device under test (DUT) for a temperature $T_{50 \Omega}$ controllable load.
		The output line is amplified by a HEMT-amplifier at 3 K  and a low noise amplifier (LNA) at room temperature. 
		The output line is connected to switch MS4 and MS3, to select between SA, VNA or down-conversion using MW LO S4 for digitizer measurement. 
		}
	\label{figSI:setup}
\end{figure}
The experimental setup consists of optics preparation, microwave preparation, CEO device in the dilution fridge, optics and microwave detection, and data acquisition.
More details of the experimental setup are shown in Fig.~\ref{figSI:setup}.

The normalized reflections of the optical modes as shown in the main text are fitted using coupled mode theory, with results shown in Table~\ref{tab:Splittings}.
The mode with largest splitting, i.e.~mode 4, is adopted as the split mode in the asymmetric case.
We note that, the obtained detunings of TE and TM mode in mode 4 are quite similar.
For this reason, we assume the same frequency for the TM mode to its corresponding TE mode in the main text.
\begin{table}
	\begin{tabular}{|l|l|l|l|l|l|l|}
		\hline
		Modes & $\kappa_o/2\pi$ (MHz) & $\kappa_{o,\mathrm{ex}}/2\pi$ (MHz) & $\delta_o/2\pi$ (MHz) & $\kappa_{o,\mathrm{tm}}/2\pi$ (MHz) & $\delta_{o,\mathrm{tm} }/2\pi$ (MHz) & $J/2\pi$ (MHz) \\ \hline
		4     & 34.6       & 8.9    & -17.8  & 7.6  & -18.5      & 26 \\ \hline
		5     & 24.7      & 9.8     & 5        & 17.4     & 28.3       & 13  \\ \hline
		6     & 24.3      & 9.2     & -3.9    & 30          & -18.7    & 10   \\ \hline
	\end{tabular}	
	\caption{\textbf{Fitted parameters for the split modes.}
		$\kappa_o$ and $\kappa_{o,\mathrm{tm}}$ correspond to the total loss rate of the TE and corresponding TM mode.
		$\delta_o$ and $\delta_{o,\mathrm{tm}}$ correspond to the cavity detuning of TE and corresponding TM mode to the main dip of the splitted mode, which is centered to zero in Fig.~\ref{fig:1}(c).}	
	\label{tab:Splittings}
\end{table}
\begin{table}
	\begin{tabular}{|l|l|l|l|l|l|}
		\hline
		Modes      & 1 and 2        & 2 and 3 & 3 and 4 & 4 and 5 & 5 and 6 \\ \hline
		Separation & 8.799GHz   & 8.799GHz   & 8.791GHz       & 8.817GHz       & 8.795GHz
		\\ \hline
	\end{tabular}
	\caption{\textbf{Calibrated frequency separation of the adjacent optical modes, shown as the distance between the main dip of each optical modes.}
	}	
	\label{tab:FSRs}
\end{table}
\begin{table}[]
	\begin{tabular}{|l|l|l|l|l|}
		\hline
		Pump Config.                       & Probing Mode                       & Pump Mode & Probe Mode & MW Frequency \\ \hline
		{\color[HTML]{35978F} Sym.}         & {\color[HTML]{BC5A5A} Stokes}      & {\color[HTML]{35978F} 2}         & 1          & 8.799 GHz    \\ \hline
		{\color[HTML]{35978F} Sym.}         & {\color[HTML]{0570B0} anti-Stokes} & {\color[HTML]{35978F} 2}         & 3          & 8.799 GHz    \\ \hline
		{\color[HTML]{BC5A5A} Stokes}      & {\color[HTML]{BC5A5A} Stokes}      & {\color[HTML]{BC5A5A} 3}         & 2          & 8.799 GHz    \\ \hline
		{\color[HTML]{0570B0} anti-Stokes} & {\color[HTML]{0570B0} anti-Stokes} & {\color[HTML]{0570B0} 5}         & 6          & 8.795 GHz    \\ \hline
	\end{tabular}
	\caption{\textbf{Measurement details for different mode and probing configurations.}
		The pump mode and probe mode indexes are given for each probing configurations.
		Microwave frequency is adjusted accordingly to match the pump and probing mode separation.	
	}
	\label{tab:probeconfig}
\end{table}
Depending on the specific pump configuration, the microwave cavity frequency is adjusted to match the optical pump and probe mode separation, as shown in Table~\ref{tab:probeconfig}.
The complete information regarding the frequency separation between the optical modes are shown in Table~\ref{tab:FSRs}.
The imperfect detunings between the Stokes and anti-Stokes modes are considered in the calculation of dynamical back-action using full theoretical model in the main text, especially regarding the optical-spring effect since it is sensitive to detunings [cf. Eq.~\ref{eq:delta_s} and Eq.~\ref{eq:delta_as}].

For the coherent response experiments in the pulsed regime, we send short optical pump pulse ($\tau\sim50\mathrm{ns}-2\mu s$) to the CEO device, while keeping the weak microwave or optical probing field on.
The optical pump pulses are triggered at rate of 100\,Hz for all the experiments, except for the Stokes case (2Hz).
We sweep the frequency around the probing mode to reconstruct the full microwave or optical response.
For each frequency, the pulses are repeated 2500 times.
In addition, we sweep the pump pulse power to investigate the power dependence of the dynamical back-action with peak power $\sim 500\mathrm{mW}$.
The RF signal from the balanced heterodyne detection of the optical probing field and the frequency down-converted microwave signal are recorded by a digitizer. 
In our experiments, both optical and microwave LO are detuned by 40MHz from the probing signal frequency.
All the dynamical back-action data are taken from the time domain traces at 1GS/s sampling rate for different mode and probing configurations, except for the delayed excess back-action data shown in Fig.~\ref{fig:4}(b) and (c) of the main text, which is taken by the SA in the zero-span mode.

\subsection{Data Analysis}
In this work, we focus on the coherent response of the multimode CEO device.
Here we show the detailed procedure for the data analysis.
\subsubsection{Susceptibility Reconstruction}
The basic principle for susceptibility reconstruction of the CEO device is shown in the main text.
The spectral normalized reflection for the probing field is defined as,
\begin{equation}
	R_j(\omega) = |S_{jj}(\omega)/S_{jj,\mathrm{off}}(\omega)|^2,
\end{equation}
where $S_{jj}(\omega)$ and $S_{jj,\mathrm{off}}(\omega)$ are the reflection coefficient of the probing field \textit{j} with pulse on and off in the lab frame.
For simplicity, we approximate the reflection with pulse off $S_{jj,\mathrm{off}}(\omega)$ to $S_{jj}(\omega)|_{t=0}$, i.e. the normalized reflection before the pulse arrives.

In the experiments, the weak coherent RF signal from the down-converted microwave and optical field $I_j(t)$ is fixed at 40MHz, more than 10 dB above the noise floor, due to the low noise amplification using HEMT amplifier or optical balanced heterodyne detection.
Here we only focus on the output power in the detection, as the phase information are washed out due to the drift between pulses.
For a phase sensitive coherent response measurement, e.g. VNA, additional weak optical pulses can be applied to obtain an insitu phase correction in each trigger.

We perform digital down-conversion (DDC) of the time-domain data at 40MHz and reconstruct the normalized reflection coefficient over time for different probe field frequencies,
\begin{equation}
	R_j(\Omega+\Omega_{\mathrm{LO},j}) = \frac{ \bar{P}_{\mathrm{out},j}(\Omega)}{ \bar{P}_{\mathrm{out},j}(\Omega)|_{t=0}},
\end{equation}
with $\Omega_{\mathrm{LO},j}$ the LO frequency and $\bar{P}_{\mathrm{out},j}$ the averaged power of the RF field from DDC.
This avoids the complicated system calibration due to the frequency dependence on the input and detection sides, especially on the optical side due to the pump filter (F2).
\subsubsection{Data Fitting of Stationary Dynamical Back-action}
As our multimode CEO device is in the normal dissipation regime, i.e.~$\kappa_o\gg\kappa_e$, microwave frequency shift and linewidth change result  in an effective susceptibility,
\begin{equation}
	\chi_{e,\mathrm{eff}}(\Omega) = \frac{1}{(\kappa_e+ \delta\kappa_e)/2 - i (\Omega - \delta\Omega_e)},
\end{equation}
where $\delta\kappa_e$ and $\delta\Omega_e$ are the linewidth and frequency change of microwave mode.
As mentioned in SI~\ref{sec:SIMethodTheory}, the on-resonance microwave probing is more susceptible to microwave frequency shift.
Considering the complex back-action dynamics, we don't adopt full model of coherent electro-optical dynamical back-action for the microwave response fitting.

On the optical side, we adopt full DBA model for the coherent response fitting, including imperfect detunings.
For the stationary dynamical back-action, we perform a joint fit of the coherent microwave and optical response at the steady regime of the pulse for all the powers, with microwave linewidth, microwave external coupling rate, optical linewidth, optical external coupling rate as shared parameters as shown in Fig.~\labelcref{figSI:1,figSI:2,figSI:3,figSI:4}.
The resulted fitting parameters, including the imperfect detunings are adopted to give the theoretical curves in Fig.~\ref{fig:2} and~\ref{fig:3} in the main text.

\begin{figure}[th]
	\includegraphics[scale=1]{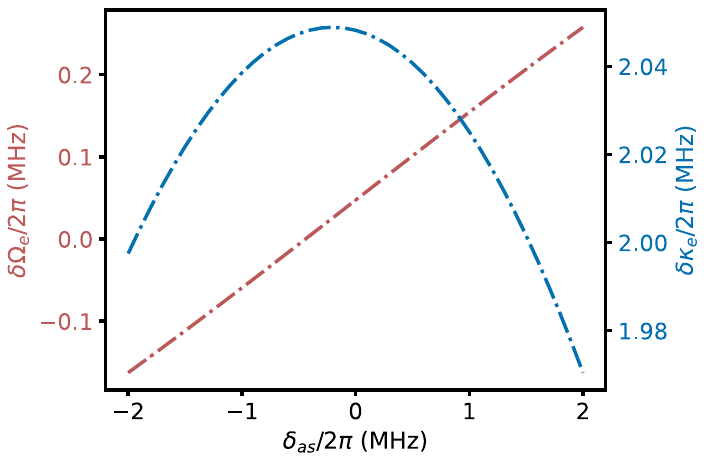}
	\caption{
		\textbf{Electro-optical back-action to the microwave mode for the anti-Stokes case.} 
		Estimated optical spring effect ($\delta\Omega_e$) and microwave linewidth change ($\delta\kappa_e$) versus the detuning uncertainty ($\delta_{\mathrm{as}} = \omega_{\mathrm{as}} - \omega_{p} -\Omega_e$) at $C=0.2$.
	}
	\label{figSI:DBAAS}
\end{figure}
In Fig.~\ref{fig:2}(b), we observe a discontinuous optical spring effect versus $C$ in the anti-Stokes case ($\omega_p = \omega_5$), especially around $C\sim0.2$.
This might be due to the detuning uncertainties in the experiments, where
we show an estimated microwave frequency and linewidth change due to imperfect detuning $\delta_{\mathrm{as}}$ at $C=0.2$ in Fig.~\ref{figSI:DBAAS}.
Minuscule optical spring effect exists for $\delta_{as}=0$, which is due to the asymmetric Stokes modes ($\omega_4$) [cf. Tab.~\ref{tab:FSRs} and~\ref{tab:Splittings}].
We note that, such detuning uncertainty is also observed in the transient dynamical back-action [cf. Fig.~\ref{figSI:InsBA}].

\subsubsection{Data Fitting of Transient Dynamical Back-action}
For the instantaneous dynamical back-action, the fitting of microwave and optical response are performed separately.
On the microwave side, for measurements with given pulse power we perform a joint fit of the coherent response over the pump pulse (from the beginning of the pulse till $\sim3\mu s$ after the pulse), to capture the delayed excess back-action due to the pump pulse.
On the optical side, we only focus on the pulsed regime, where $C$ over time is obtained as shown in Fig.~\ref{figSI:InsBA}(a), since the optical coherent response restores instantaneously after the pulse off. 
We note that, to capture perfectly the temporal dynamics on the optical side, \textit{a free parameter $\delta = \Omega_e - \mathrm{FSR}$ needs to be introduced in the fitting}, which indicates pump pulse induced FSR or microwave frequency change during the pulse, as explained in the next section (SI~\ref{sec:INEXBA}).
The resulted fitting parameters are adopted to give the theoretical curves during the pulse in Fig.~\ref{fig:4}(a). 
The obtained microwave frequency and linewidth change after the pulse are shown in Fig.~\ref{figSI:DelayBA}.

\subsection{Instantaneous Excess Back-action}\label{sec:INEXBA}
\begin{figure}[th]
	\includegraphics[scale=1]{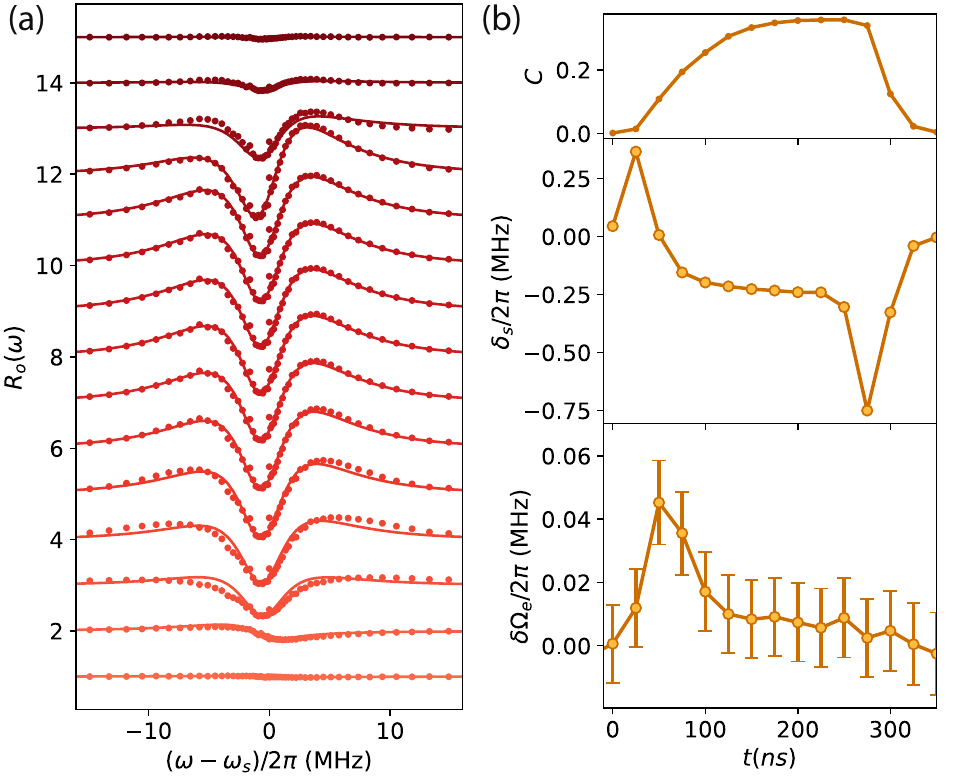}
	\caption{
		\textbf{Instantaneous excess back-action over the pulse for the symmetric case.} 
	\textbf{a,} Displaced instantaneous optical normalized reflection $R_o(\omega)$ around the Stokes mode during the pulse, with corresponding fitting curves. 
	 \textbf{b,} Upper panel, fitted $C$ from optical response over the pulse. 
	 Middle panel, fitted $\delta_s$, i.e.~$ \Omega_e - \mathrm{FSR}$ from optical response over the pulse.
	  Lower Panel,  fitted $\delta\Omega_e$ from microwave response over the pulse. Error bars indicate two standard deviations.
	}
	\label{figSI:InsBA}
\end{figure}
In Fig.~\ref{figSI:InsBA}(a), we show the displaced coherent optical response $R_o(\omega)$ over the pulse with corresponding fitting curves. 
During the loading and unloading of the optical pump, $R_o(\omega)$ is not symmetric around the Stokes mode resonance, which indicates frequency mismatch between $\Omega_e$ and the optical mode separation, i.e.~$\delta_s = \Omega_e -\mathrm{FSR}\neq 0$.
In Fig.~\ref{figSI:InsBA}(b), we show the fitted $C$ and $\delta_s$ in the upper and middle panel, while $\delta\Omega_e$ in the lower panel.
We note that, a detuning change of $\sim1\mathrm{MHz}$ arises during the loading and unloading of the pulse.
The exact reason for the detuning change requires further exploration.
It can be attributed to either optical FSR change, e.g. due to photo-refractive effect~\cite{xu_photorefractioninduced_2021} or dissipative feedback~\cite{qiu_dissipative_2022}, or intrinsic microwave frequency change, e.g. due to quasi-particles~\cite{witmer_siliconorganic_2020,mirhosseini_superconducting_2020}.

\subsection{Excess Back-action}\label{sec:ExDelayedBA}
\begin{figure}[th]
	\includegraphics[scale=1]{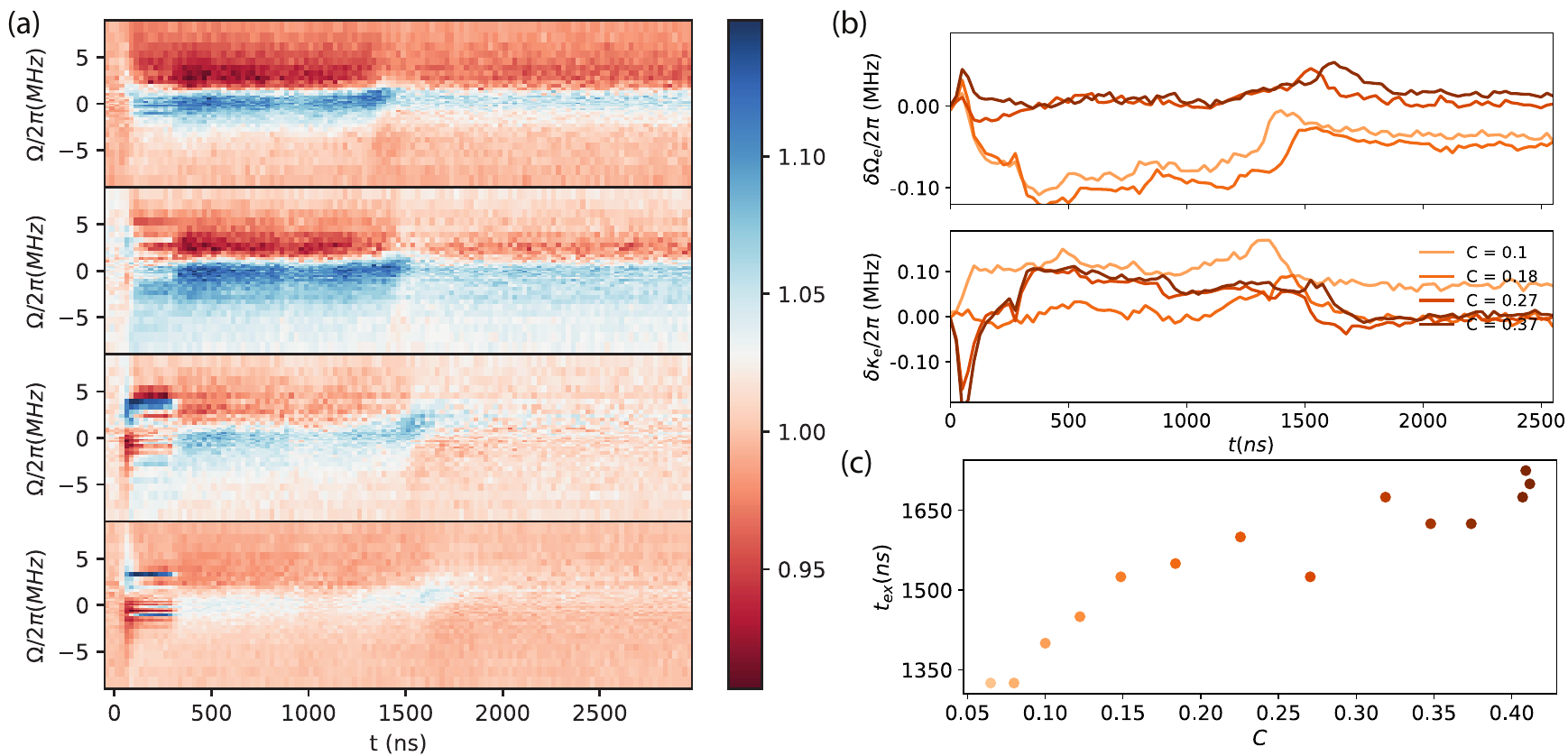}
	\caption{
		\textbf{Excess back-action to the microwave mode for the symmetric case.} 
		\textbf{a}, Normalized reflection coefficients $R_e(\Omega)$ over the pump pulse (from the top to bottom) for $C$ of 0.1, 0.18, 0.27, and 0.37. 
		\textbf{b}, Fitted microwave frequency and linewidth change after the pulse from (a).
		Different color corresponds to different $C$.
		\textbf{c}, Extracted $C$ dependent $t_{\mathrm{ex}}$, i.e.~the transition time for the excess back-action after the pulse is off.
		}
	\label{figSI:DelayBA}
\end{figure}
In Fig.~\ref{fig:4}(b) of the main text, we show the excess delayed back-action after the pulse with resonant ($\Omega_e = \mathrm{FSR}$) and off-resonant ($\Omega_e \neq \mathrm{FSR}$) comparison, for the symmetric mode configuration ($J_{\mathrm{s/as}}=0$).  
In Fig.~\ref{figSI:DelayBA}, we show the excess back-action to the microwave mode of the symmetric case in Fig.~\ref{fig:4}(a), with extracted microwave frequency and linewidth change.
Figure~\ref{figSI:DelayBA}(a) shows the 2D plot of $R_e(\Omega)$ during the pulse,
where the large color contrast around resonance indicates microwave frequency shift [cf. Eq.~\ref{eq:SIRonres}].
Even for low $C$, the excess delayed back-action is rather evident after the pulse is off,
where we observe a dramatic microwave frequency shift at $t_{\mathrm{ex}}\sim1300ns$, despite of the absent optical pump pulse. 
Figure~\ref{figSI:DelayBA}(b) shows the fitted frequency and linewidth change during the pulse.
We note that, for high C, the microwave linewidth decreases when the pulse arrives, which is consistent with  Fig.~\ref{fig:4}(a) (middle panel) and the microwave reflection decrease in Fig.~\ref{fig:1}(e) (upper left). 
As $C$ increases, the color contrast decreases, while $t_{\mathrm{ex}}$ slowly increases, which is consistent with the results in Fig.~\ref{fig:4}(b).
The exact underlying dynamics remains further exploration, which might be related to the laser induced quasi-particles in the microwave cavity~\cite{witmer_siliconorganic_2020, mirhosseini_superconducting_2020}.
\clearpage
\begin{figure*}[ht!]
	\includegraphics[scale=0.7]{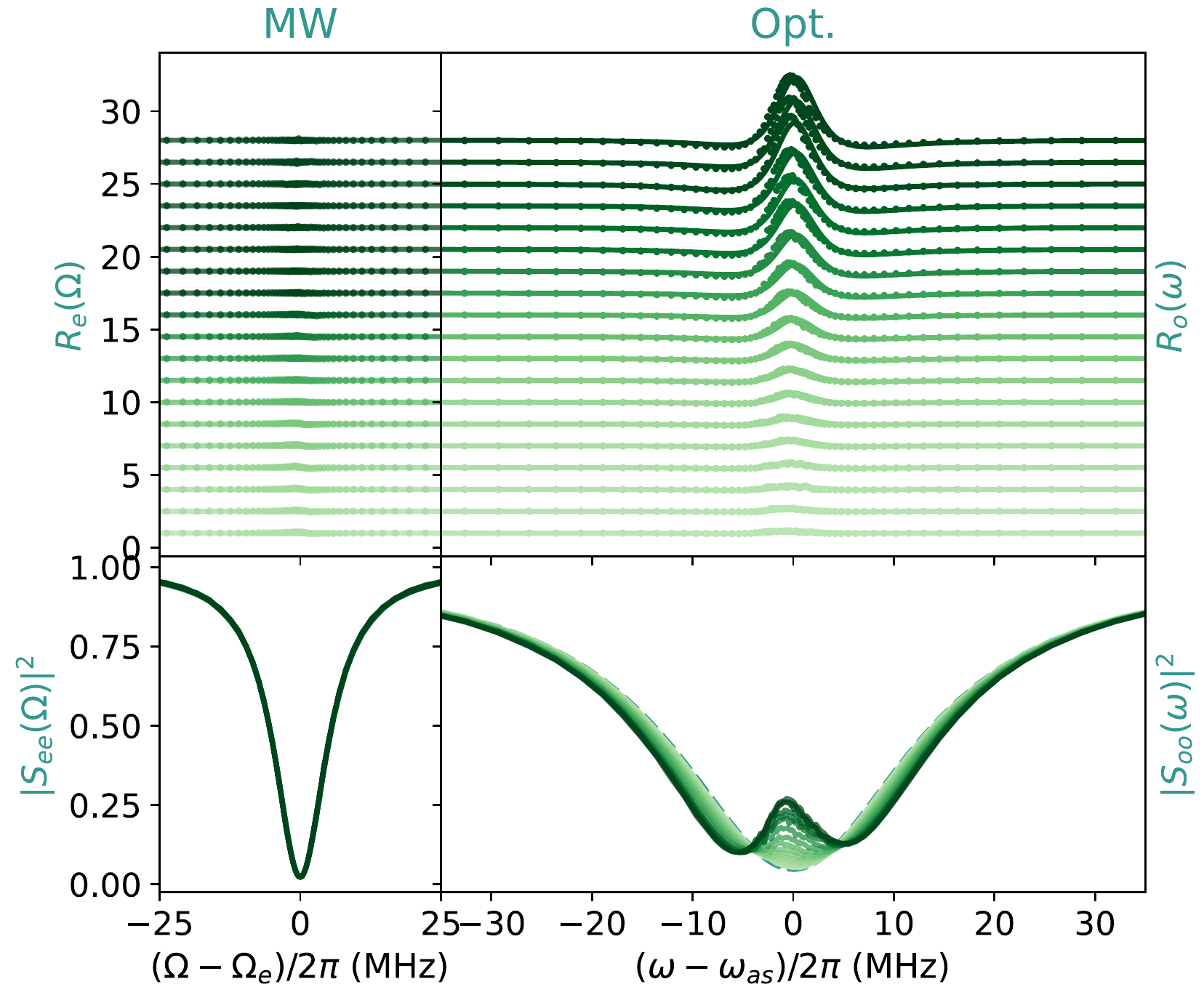}
	\caption{\textbf{Coherent stationary response of microwave and anti-Stokes mode in the symmetric mode configuration ($J_{\mathrm{s/as}}=0$)   at different $C$ as in Fig.~\labelcref{fig:2,fig:3}}.
		\label{figSI:1}}
\end{figure*}

\begin{figure*}[ht!]
	\includegraphics[scale=0.7]{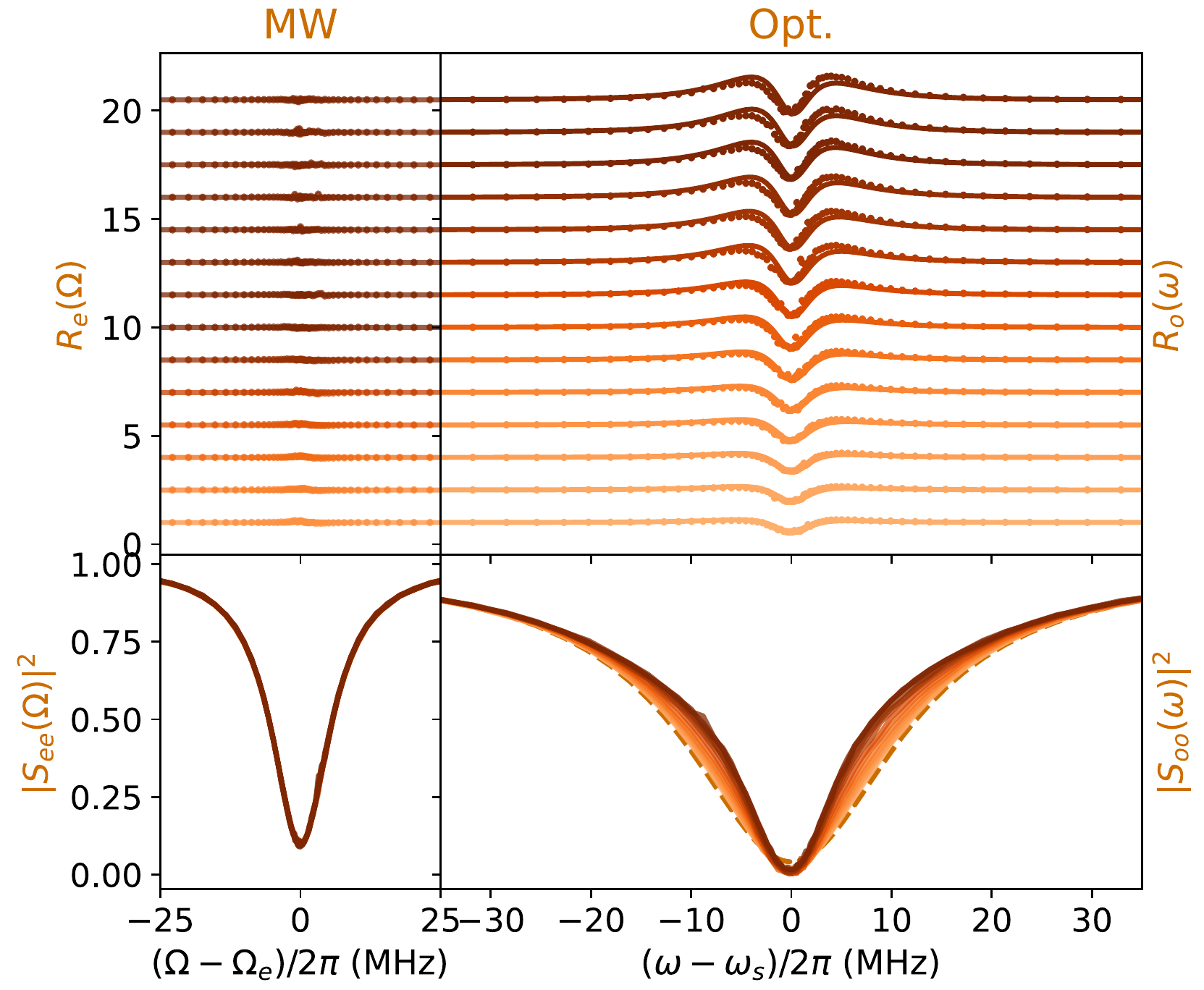}
	\caption{\textbf{Coherent stationary response of microwave and optical Stokes mode in the symmetric mode configuration ($J_{\mathrm{s/as}}=0$) at different $C$ as in Fig.~\labelcref{fig:2,fig:3}}.
		\label{figSI:2}}
\end{figure*}

\begin{figure*}[ht!]
	\includegraphics[scale=0.7]{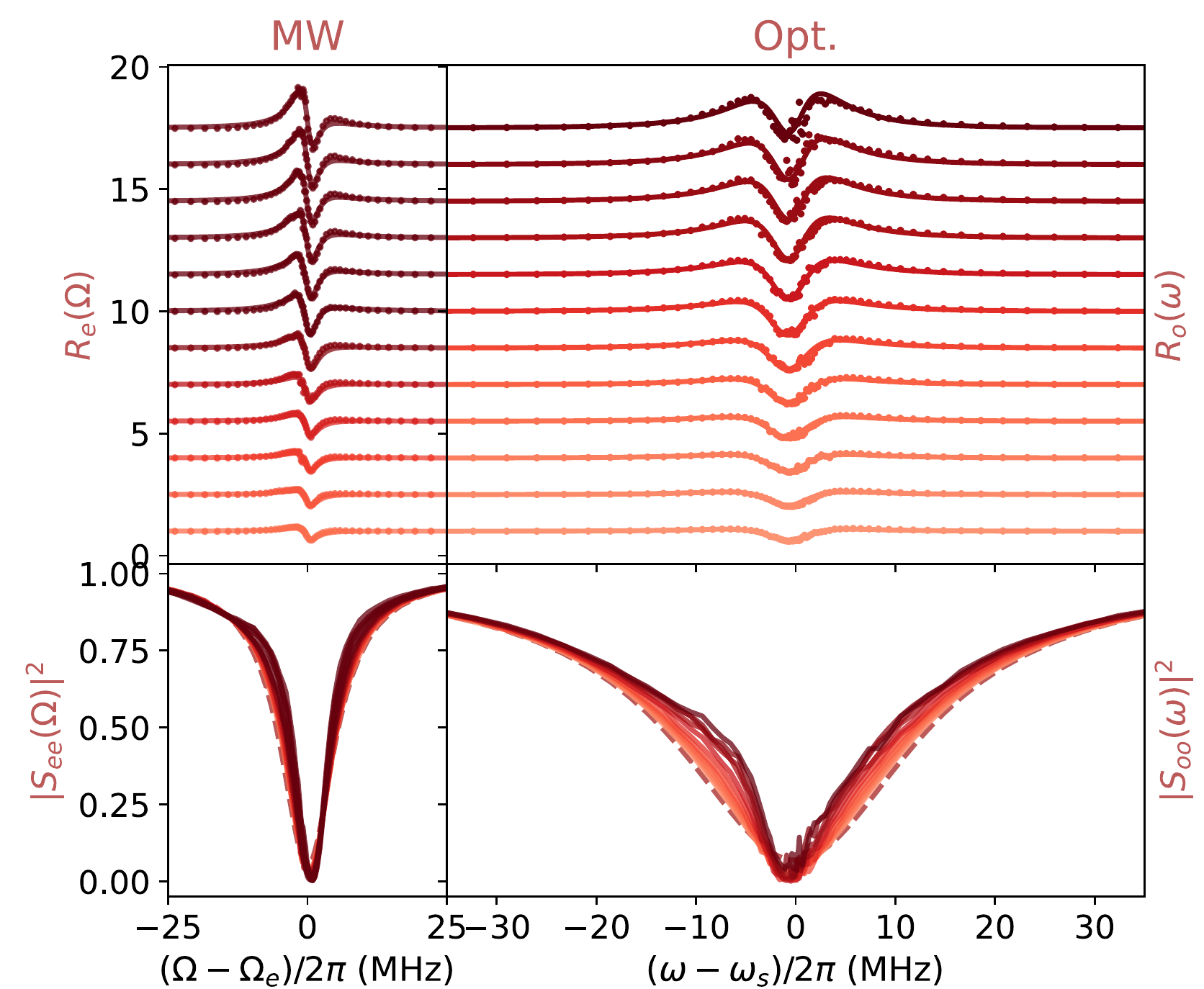}
	\caption{\textbf{Coherent stationary response of microwave and Stokes mode in the Stokes case ($J_{\mathrm{s}}=0$)   at different $C$ as in Fig.~\labelcref{fig:2,fig:3} }.
		\label{figSI:3}}
\end{figure*}

\begin{figure*}[ht!]
	\includegraphics[scale=0.7]{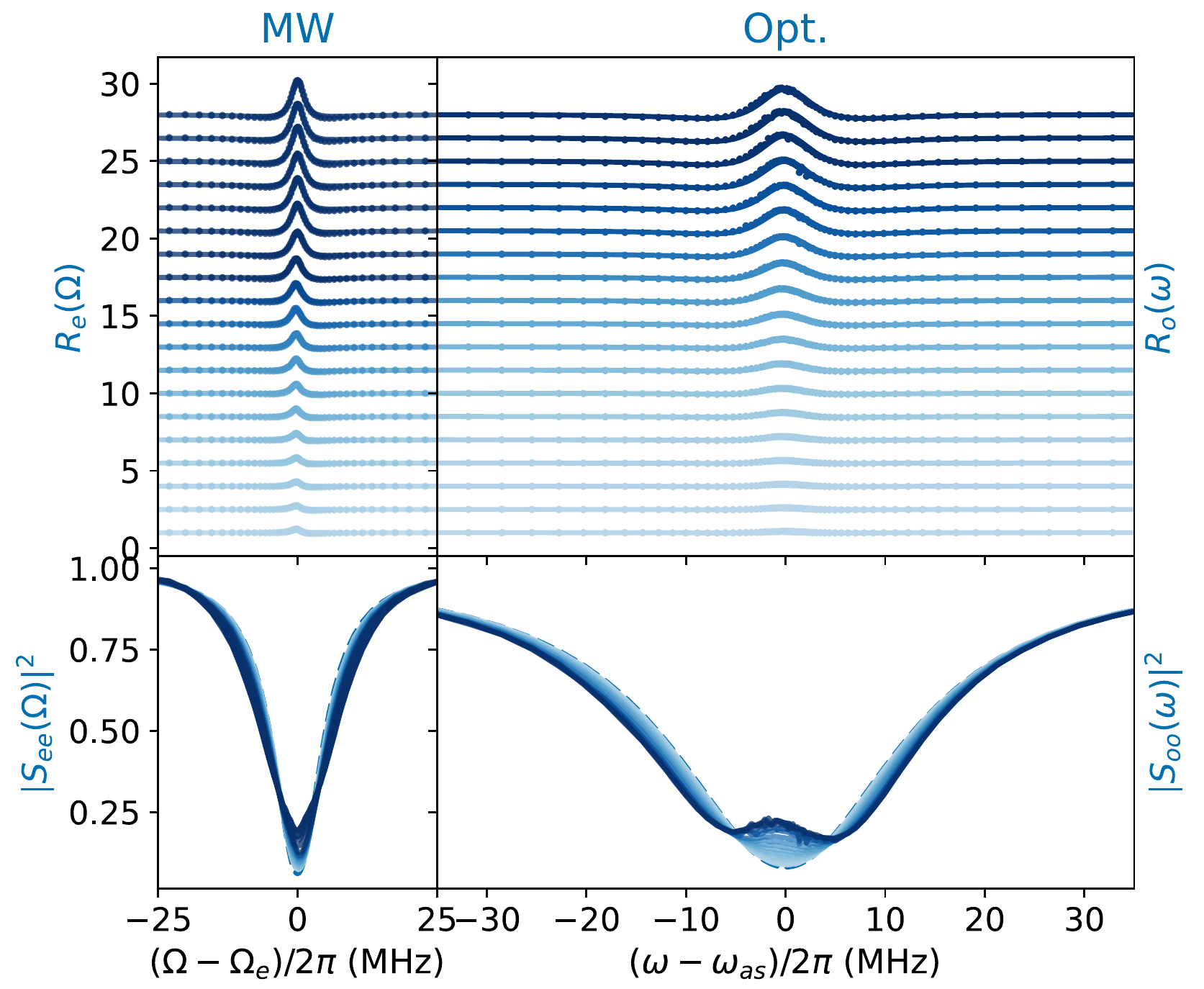}
	\caption{\textbf{
		Coherent stationary response of microwave and anti-Stokes mode in the anti-Stokes case ($J_{\mathrm{as}}=0$)  at different $C$ as in Fig.~\labelcref{fig:2,fig:3}}.
		\label{figSI:4}}
\end{figure*}

\begin{figure*}[ht!]
	\includegraphics[scale=0.7]{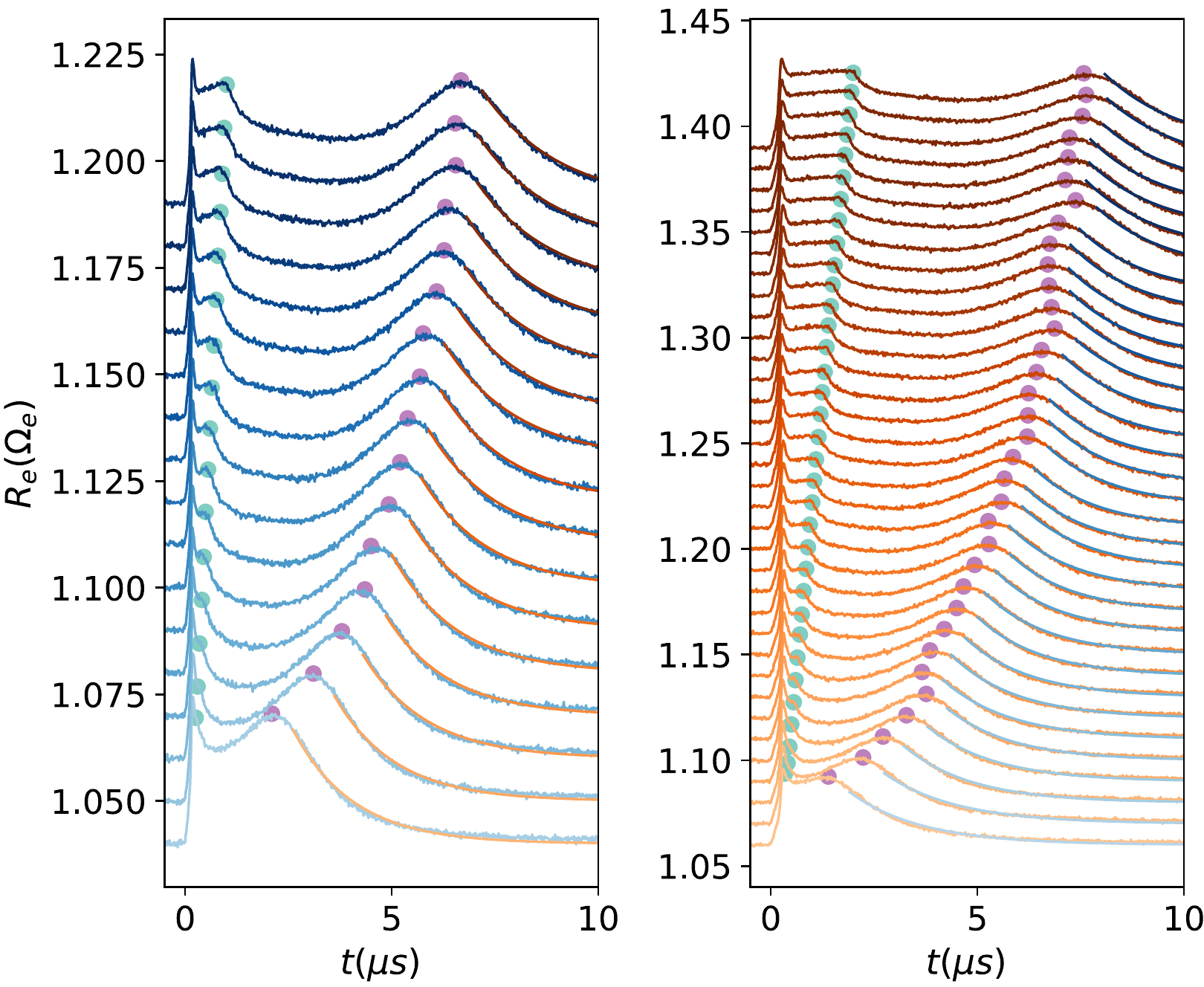}
	\caption{\textbf{
			Instantaneous coherent microwave response in the symmetric case ($J_{\mathrm{s/as}}=0$) with different pulse length as in Fig.~\ref{fig:4}(c). 
			}
		The resonant ($\Omega_e = FSR$) and off-resonant ($\Omega_e \neq FSR$) cases are shown in the left and right panels.
		The corresponding pulse end and the bounce of the delayed back-action are indicated with green and purple dots, where the time difference is $t_{\mathrm{ex}}$. The mean lifetime $\tau_{\mathrm{ex}}$ is fitted from exponential decay of $R_e(\Omega_e)$ after the bounce, shown as orange and blue curves in the left and right panels. 
		\label{figSI:5}}
\end{figure*}
\end{document}